\documentclass[floatfix,superscriptaddress,numerical,amsmath,amssymb, pre, aps, reprint,showpacs]{revtex4-1}

\usepackage{graphicx}
\usepackage{dcolumn}
\usepackage{float}
\usepackage{hyperref}
\usepackage{subfigure}

\begin{document}

\title{A Heterogeneous Out-of-Equilibrium Nonlinear $q$-Voter Model with
Zealotry}
\author{Andrew Mellor}
\affiliation{Department of Applied Mathematics, School of Mathematics, University of
Leeds, Leeds LS2 9JT, U.K.}
\author{Mauro Mobilia}
\affiliation{Department of Applied Mathematics, School of Mathematics, University of
Leeds, Leeds LS2 9JT, U.K.}
\author{R.K.P. Zia}
\affiliation{Center for Soft Matter and Biological Physics, Department of
Physics, Virginia Polytechnic Institute \& State University, Blacksburg, VA
24061, USA}

\begin{abstract}
We study the dynamics of the out-of-equilibrium nonlinear $q$-voter model
with two types of susceptible voters and zealots, introduced in [EPL \textbf{%
113}, 48001 (2016)]. In this model, each individual supports one of two
parties and is either a susceptible voter of type $q_1$ or $q_2$, or is an
inflexible zealot. At each time step, a $q_i$-susceptible voter ($i=1,2$)
consults a group of $q_i$ neighbors and adopts their opinion if all group
members agree, while zealots are inflexible and never change their opinion.
This model violates detailed balance whenever $q_1 \neq q_2$ and is
characterized by two distinct regimes of low and high density of zealotry.
Here, by combining analytical and numerical methods, we investigate the
non-equilibrium stationary state of the system in terms of its probability
distribution, non-vanishing currents and unequal-time two-point correlation
functions. We also study the switching time properties of the model by
exploiting an approximate mapping onto the model of [Phys. Rev. E \textbf{92},
012803 (2015)] that satisfies the detailed balance, and we outline some properties of the model near criticality.
\end{abstract}

\maketitle

\section{Introduction}

\label{sec:introduction} Parsimonious individual-based models have been
commonly used to describe collective social phenomena for more than four
decades~\cite{schelling1969models,*schelling1971dynamic,*schelling1978micromotives}. In particular, statistical physics models have
proven especially well-suited to reveal the micro-macro connections in
social dynamics and to help reveal the relationships existing between
phenomena appearing across disciplines (like biology, ecology, economics)~%
\cite{castellano2009statistical}. In particular, the voter model (VM)~\cite{liggett1985interacting} serves
as a reference to describe the evolution of opinions in interacting
populations~\cite{castellano2009statistical,galam2012sociophysics,*sen2013sociophysics}. It is however well established
that the VM is built on a number of oversimplified assumptions: For
instance, the VM considers that all voters are similar: In the absence of
any self-confidence, they change their their opinion by imitating neighbors.
This mechanism of \textit{conformity by imitation} does not allow us to
maintain social diversity in the long run, but always leads to a consensus.
However, social scientists have shown that the collective dynamics of a
society greatly depends on the different responses to stimuli by the members
of a society~\cite{granovetter1978threshold,latane1981psychology,*nail2000proposal,asch1955opinions,*milgram1969note}. A simple way to
accommodate a population with different levels of confidence is to
assume that some agents are ``zealots'' and either favor one opinion~\cite{mobilia2003does,*mobilia2005voting} or inflexibly maintain a fixed opinion~\cite{mobilia2007role}. Since the
introduction of zealots in the VM, the effect of zealotry has been studied
in other models of social dynamics~\cite{galam2007role,*sznajdweron2011phase,*acemoglu2013opinion,*nyczka2013anticonformity,*yildiz2013binary,*palombi2014stochastic,*verma2014impact,*waagen2015effect,*arendt2015opinions} and in various
contexts~\cite{masuda2015opinion,*xie2011social,*masuda2012evolution,*borile2015coexistence,*mobilia2012stochastic,*mobilia2013evolutionary,
*mobilia2013reply,*mobilia2013commitment,*szolnoki2016zealots}. 

In this work, we focus on the so-called two-state nonlinear $q$-voter model ($q$VM)~\cite{castellano2009nonlinear}. In this variant of the VM, that has received much
attention recently~\cite{slanina2008some,*galam2011pitfalls,*przybyla2011exit,*timpanaro2014exit,
*timpanaro2015analytical,*jedrzejewski2015oscillating,*javarone2015conformism}, each voter can be influenced by a group
of $q$ neighbors~\cite{[{The version with $q=2$ is closely related to the well-known models of }][{}]sznajdweron2000opinion,
*slanina2003analytical,*lambiotte2008dynamics}
~\footnote{
In this work, we will ignore spatial structures, i.e., each voter is
connected to every other, so that the term \textquotedblleft
neighbor\textquotedblright\ simply refers to any other individual in the
population. Of course, to model reality more closely, we should study voters
linked in more complex networks~\cite{chuang2016bistable}.}. This mimics the fact that group pressure is known
to influence the degree of conformity, especially above a group size
threshold~\cite{asch1955opinions,*milgram1969note}. Furthermore, social scientists have shown that
conformity via imitation is a driving mechanism for collective actions that is influenced by 
the size of social groups, and they also found that conformity can
be seriously deflected by individuals (like zealots) that are able to resist
 group pressure~\cite{asch1955opinions,*milgram1969note,latane1981psychology,*nail2000proposal}. Motivated by these
considerations~\cite{latane1981psychology,*nail2000proposal,granovetter1978threshold}, 
the basic features of the $q$VM and zealotry have recently been combined in
the $q$-voter model with inflexible zealots ($q$VMZ)~\cite{mobilia2015nonlinear} and in a
heterogeneous counterpart model, called the 2$q$VZ, in which two subgroups
of susceptible voters interacting with their neighbors with different $q$'s:
namely, with $q_{1}<q_{2}$~\cite{mellor2016characterization}.

In the $q$VMZ and 2$q$VZ, group-size limited conformity and zealotry are
accounted for, and zealots significantly tame the level of social conformity
in the population~\cite{mobilia2015nonlinear,mellor2016characterization}. Furthermore, the dynamics of the $q$VMZ
and 2$q$VZ in a well-mixed population is similar and is characterized by two
phases as in systems in thermal equilibrium: Below a critical level of
zealotry, the opinion distribution transitions from single-peaked becomes
bimodal and, in finite populations, the dynamics is characterized by the
fluctuation-driven switching between two states~\cite{mobilia2015nonlinear,mellor2016characterization}. Yet,
while the $q$VMZ of Ref.~\cite{mobilia2015nonlinear} obeys detailed balance and its 
stationary distribution can be obtained exactly, this is not the case of the
2$q$VZ even in the simple case of a well-mixed population when
the system relaxes into a non-equilibrium steady state (NESS)~\cite{mellor2016characterization}. 
The fact that the 2$q$VZ does
not obey detailed balance has many important statistical physics
consequences: There is generally no simple way to obtain the NESS
distribution~\cite{hill1966studies}, while persistent probability currents are
responsible for subtle oscillations and non-trivial correlation functions~%
\cite{zia2007probability,mellor2016characterization}. From a social dynamics viewpoint, the composition of the
society being very heterogeneous, it would be relevant to consider models
with a distribution of $q$'s that would reflect different responses to
multiple social stimuli~\cite{granovetter1978threshold}. While analytical progress
appears to be difficult in such a general case, much can be learnt about the
influence of being out-of-equilibrium on the social dynamics by considering
the simple 2$q$VZ whose detailed analysis is presented here. In particular,
we are able to show that susceptible voters with smaller $q$ change their
opinion more quickly and drive those in the other subgroup, and directed,
yet subtle, \textit{oscillations} associated with fluctuating quantities can
be measured~\cite{[{}][{ (See also arXiv:1610.02976)}]zia2016manifest}. 

In this work, we therefore focus on the 2$q$VZ and characterize the
properties of its NESS in terms of the master equation, stochastic
simulations and continuum approximations based on mean-field and
Fokker-Planck equations. While  a brief account of some of these features was given in Ref.~\cite{mellor2016characterization},
here  we  considerably generalize that study and  also investigate
the switching dynamics of the 2$q$VZ.

The remainder of this paper is organized as follows: The details of the 2$q$VZ are specified in the next section, along with the underlying master
equation. The microscopic characterization of the NESS in terms of the
master equation and relevant observables is addressed in Sec.~III. Section IV
is dedicated to the continuum approximation  of the $2q$VZ dynamics in terms of mean-field rate equations and through a
linear Gaussian approximation (LGA) of the underlying Fokker-Planck equation (for a system of large but finite size)~
\cite{lax1966classical,weiss2003coordinate, *weiss2007fluctuation,zia2007probability}. This allows us to
show in some
details, how the LGA offers the best insight into probability currents
and their manifestations in the $2q$VZ's NESS far from criticality, as
explained in Sec.~IV.B. In Sec.~IV.C, we exploit a Fokker-Planck formulation
to shed light on the switching time properties of the 2$q$VZ via an approximate mapping onto the $q$VMZ whereas the behavior near criticality is briefly
discussed in Sec.~IV.D. While most examples given in Sec.~IV are
concerned with the case of symmetric zealotry (the same number of zealots for
each party), the generalization to asymmetric zealotry is addressed in
Sec.~V. We also give a  number of analytical and computational details  in a
series of appendixes. We end with a summary, as well as an outlook for future research.
It is worth noting that while our general analysis applies to any values of 
$q_1$ and $q_2$, for the sake of simplicity and without loss of generality, 
the numerical examples reported in our figures
have been obtained for $q_{1,2}=1,2$.

\section{The 2$q$ Voter Model with Zealotry (2$q$VZ)}

\label{sec:model}

\begin{figure}[tbp]
\centering
\subfigure[]{\fbox{\includegraphics[width=0.29%
\linewidth]{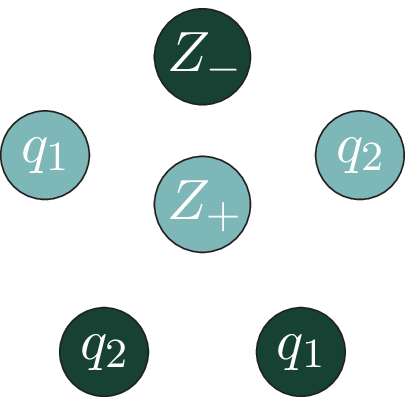}}} 
\subfigure[]{\fbox{\includegraphics[width=0.29%
\linewidth]{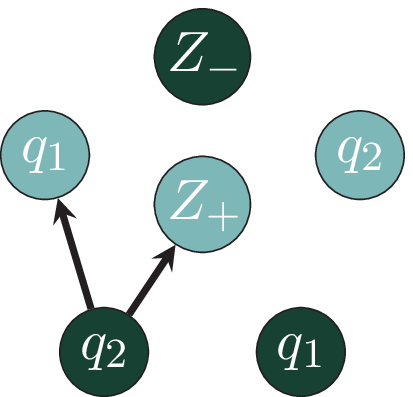}}} 
\subfigure[]{\fbox{\includegraphics[width=0.29%
\linewidth]{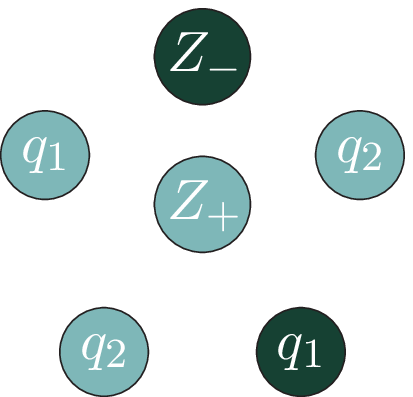}}}
\caption{\textit{(Color online)}. Illustration of the 2$q$VZ dynamics: (a)
the system consists of zealots, $q_{1}$-, and $q_{2}$- susceptibles; each
can hold one of two opinions (green/black or blue/grey). (b) A $q_{i}$ susceptible
picks $q_{i}$ neighbors at random. Here, $q_{2}=2$. (c) If the picked
neighbors have same opinion, the focal susceptible voter adopts their
opinion, otherwise no change occurs.}
\label{fig:model}
\end{figure}
The 2$q$VZ consists of a population of $N$ voters who hold one of two
opinions, denoted by $\pm 1$. A fraction of the population are inflexible
zealots that never change their opinion~\cite{mobilia2007role,mobilia2015nonlinear}, the number of
which are denoted by $Z_{\pm }$. The remaining population  consists of $S=S_1+S_2$ swing voters
of two types, $q_{1}$ and $q_{2}$ (known as $q_{i}$-susceptibles, $i=1,2$): $S_{1}$ of the
swing voters are of type $q_1$ and there are $S_2$ swing voters  of  type $q_2$. In our model, the
behavior of each voter is fixed, so that $Z_{\pm }$ and $S_{i}$ are
conserved, with $S_{1}+S_{2}+Z_{+}+Z_{-}=N$. As illustrated in Fig.~\ref%
{fig:model}, the evolution of the 2$q$VZ is such that at each update attempt
(a) a (focal) voter is chosen at random. (b) If the chosen voter is a zealot
then no further action is taken. If the focal voter is a $q_{i}$-susceptible however, then the opinions of a random group of $q_{i}$ of its
neighbors (repetition is allowed) are collected. (c) If the $q_{i}$
neighbors have the same opinion that differs from that of the focal voter,
the latter adopts the opinion of the group. If there is no consensus within
the $q_{i}$-group, the opinion of the focal voter does not change. Here, 
for the sake of simplicity we consider a well-mixed population, i.e., each voter can be thought of as one of the $N$ nodes of a
complete graph and all other voters are \textquotedblleft
neighbors\textquotedblright.

\subsection{Microscopic Description with a Master Equation}

In the absence of explicit spatial structure, the configuration space is a
discrete set of $S_{1}\times S_{2}$ points and the system state is
completely specified by the pair $\vec{n}:=(n_{1},n_{2})$ where $n_{i}$ is
the number of susceptible voters of type $i$ holding the opinion $+1$. In
each update attempt, the system may, or may not, step to one of its four
nearest neighboring points. In other words the allowed \textit{changes} are
in the set $\left\{ \pm \vec{e}_{1},\pm \vec{e}_{2}\right\} $, with $\vec{e}%
_{1}:=(1,0)$ and $\vec{e}_{2}:=(0,1)$. Thus, we may regard the evolution of
the 2$q$VZ as a two-dimensional random walker with inhomogeneous and biased
rates. Such a stochastic process can be readily described by a master
equation (ME) for the evolution of $P(\vec{n},T)$, the probability of
finding the system in state $\vec{n}$ after $T$ discrete time steps (or
update attempts) from starting in an initial configuration $\vec{n}_{0}$.
One of our main interests is in the unique stationary distribution, which
has no dependence on the initial configuration $\vec{n}_{0}$, and we can
therefore omit any reference to $\vec{n}_{0}$.

For any discrete Markov process, the ME can be written as%
\begin{equation}
P(\vec{n}^{\prime },T+1)=\sum_{\vec{n}}\mathcal{G}(\vec{n}^{\prime },\vec{n}%
)P(\vec{n},T)  \label{eqn:ME}
\end{equation}%
where $\mathcal{G}$ is known as the transition matrix, or evolution
operator. In our case, since $\vec{n}^{\prime }\in \{\vec{n},\vec{n}\pm \vec{%
e}_{1},\vec{n}\pm \vec{e}_{2}\}$, so that, an explicit form for $\mathcal{G}$
is

\begin{eqnarray}
\mathcal{G}(\vec{n}^{\prime },\vec{n}) &=&\delta \left( n_{1}^{\prime
},n_{1}\right) \delta \left( n_{2}^{\prime },n_{2}\right) W^{0}(\vec{n})
\label{Gexpl} \\
&+&\sum_{i=1,2,j\neq i}\delta \left( n_{i}^{\prime },n_{i}+1\right) \delta
\left( n_{j}^{\prime },n_{j}\right) W_{i}^{+}(\vec{n})  \notag \\
&+&\sum_{i=1,2,j\neq i}\delta \left( n_{i}^{\prime },n_{i}-1\right) \delta
\left( n_{j}^{\prime },n_{j}\right) W_{i}^{-}(\vec{n}).  \notag
\end{eqnarray}%
Here, $W^{0}(\vec{n})$ represents the probability for the system to remain
unchanged, while $W_{1}^{\pm }(\vec{n})$ and $W_{2}^{\pm }(\vec{n})$ stand
for the stepping probabilities associated with $\vec{n}\rightarrow \vec{n}%
\pm \vec{e}_{1}$ and $\vec{n}\rightarrow \vec{n}\pm \vec{e}_{2}$ ,
respectively. Explicitly~\footnote{We have found the following typos in our Ref.~[17]: Eq.~(2) of [17] should
read $W_{i}^{-}(\vec{n})=\frac{n_i}{N}\left(\frac{Z_- +S_1 +S_2 -n_1-n_2}{N-1}\right)^{q_i}$,
and  after Eq.~(3) of [17], the quantity $\rho$ should read $(1-\mu^*)/\mu^*$.
}, these are%
\begin{eqnarray}
W_{i}^{+}(\vec{n}) &=&\frac{S_{i}-n_{i}}{N}\left( \frac{M}{N-1}\right)
^{q_{i}}  \notag  \label{W} \\
W_{i}^{-}(\vec{n}) &=&\frac{n_{i}}{N}\left( \frac{N-M}{N-1}\right) ^{q_{i}}
\label{eqn:w} \\
W^{0}(\vec{n}) &=&1-W_{1}^{+}(\vec{n})-W_{1}^{-}(\vec{n})-W_{2}^{+}(\vec{n}%
)-W_{2}^{-}(\vec{n}),  \notag
\end{eqnarray}%
where $M=Z_{+}+n_{1}+n_{2}$ is the total number of $+1$ voters before the
update. (See (\ref{evolG}) in Appendix~\ref{app:AppendixA} for further
details.)

Subsequently, we will be interested in the joint probability, 
\begin{equation}
\mathcal{P}(\vec{n}^{\prime },T^{\prime };\vec{n},T)=\mathcal{G}^{T^{\prime
}-T}(\vec{n}^{\prime },\vec{n})P(\vec{n},T),  \label{eqn:joint}
\end{equation}%
for finding the system being in a state $\vec{n}^{\prime }$ at (a later,
assuming $T^{\prime }>T$) time $T^{\prime }$ \textit{and} being in state $%
\vec{n}$ at time $T$ (see (\ref{jointP})). With these probability
distributions, we can compute physical observables such as the average
number of $q_{i}$ voters holding opinion $+1$ at time $T$ and two-point
correlation functions at general times: 
\begin{eqnarray*}
\langle n_{i}\rangle _{T} &=&\sum_{\vec{n}}n_{i}P(\vec{n},T) \\
\langle n_{i}^{\prime }n_{j}\rangle _{T^{\prime },T} &=&\sum_{\vec{n},\vec{n}%
^{\prime }}n_{i}^{\prime }n_{j}\mathcal{P}(\vec{n}^{\prime },T^{\prime };%
\vec{n},T).
\end{eqnarray*}%
One quantity of particular interest for the study of the 2$q$VZ is the \emph{net probability current}, 
given by $\vec{K}(\vec{n};T) = (K_1, K_2)$, see (\ref{curr}) in Appendix~\ref{app:AppendixA}. Here, $K_i(\vec{n}; T) = W_i^+(\vec{n})P(\vec{n};T) - 
W_i^-(\vec{n}')P(\vec{n}'; T)$  denotes the net flow (of probability) from $\vec{n}$ to 
$\vec{n}' \equiv \vec{n} + \vec{e_i}$, with $i=1,2$,  $\vec{e}_1=(1,0)$ and $\vec{e}_2=(0, 1)$.

\subsection{Violation of Detailed Balance}

\begin{figure}[tbp]
\centering
\includegraphics[width=0.7\linewidth]{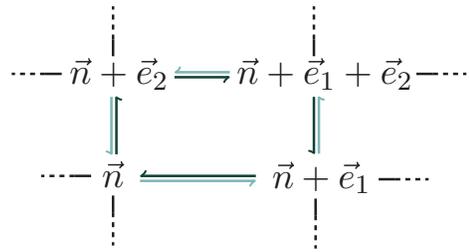}
\caption{\textit{(Color online)}. The state space of the 2$q$VZ is of size $%
(S_{1}+1)\times (S_{2}+1)$ and can be represented as a two-dimensional
lattice. The probability of traversing the shown loop clockwise (green/black) does
not equal the probability of traversal in the opposite direction (blue/grey). By
Kolmogorov's criterion, the detailed balance is violated; see text.}
\label{fig:detailedbalance}
\end{figure}

In Ref.~\cite{mellor2016characterization}, we emphasized a distinctive feature of the 2$q$VZ,
namely, that its dynamics is a genuine out-of-equilibrium type, contrary to
the $q$VMZ~\cite{mobilia2015nonlinear} model. Hence, detailed balance and time reversal symmetry are
violated in the 2$q$VZ. To establish this fact, we apply the Kolmogorov criterion (%
\ref{kol})~\cite{kolmogorov1936zur} on a closed loop consisting of the four $\vec{n}$'s
around a square: $\vec{n}\rightarrow \vec{n}+\vec{e}_{1}\rightarrow \vec{n}+%
\vec{e}_{1}+\vec{e}_{2}\rightarrow \vec{n}+\vec{e}_{2}\rightarrow \vec{n}$,
illustrated in Fig.~\ref{fig:detailedbalance}. The product of transition
probabilities around this loop is 
\begin{equation*}
W_{1}^{+}\left( \vec{n}\right) W_{2}^{+}\left( \vec{n}+\vec{e}_{1}\right)
W_{1}^{-}\left( \vec{n}+\vec{e}_{1}+\vec{e}_{2}\right) W_{2}^{-}\left( \vec{n%
}+\vec{e}_{2}\right) .
\end{equation*}%
Meanwhile, the product for the \textit{reverse} loop is 
\begin{equation*}
W_{2}^{+}\left( \vec{n}\right) W_{1}^{+}\left( \vec{n}+\vec{e}_{2}\right)
W_{2}^{-}\left( \vec{n}+\vec{e}_{1}+\vec{e}_{2}\right) W_{1}^{-}\left( \vec{n%
}+\vec{e}_{1}\right)
\end{equation*}%
so that the ratio of the two probabilities is%
\begin{equation*}
\left( \frac{M+1}{M}\frac{N-M-1}{N-M-2}\right) ^{q_{1}-q_{2}}.
\end{equation*}%
Since the quantity in the bracket is strictly greater than unity, this ratio
is not unity as long as $q_{1}\neq q_{2}$. Thus, the dynamics of our 2$q$VZ
violates the detailed balance. Of course, setting $q_{1}=q_{2}$  we recover
the $q$VMZ, for which the above ratio is always unity, confirming that the single-%
$q$ case of Ref.~\cite{mobilia2015nonlinear} satisfies the detailed balance. 

Violation of detailed balance has a number of consequences on the behavior
of a system. These will be scrutinized in the next sections. In particular,
we will be mostly interested in the behavior of the stationary state
(reached after very long times from any initial $\vec{n}_{0}$), described by
the $T$-independent distribution $P^{\ast}\left(\vec{n}\right)$~\footnote{Henceforth,
 we  denote quantities associated with the stationary state with $^{\ast
}$, unless explicitly stated otherwise.}.
From Eq.~(\ref{eqn:ME}), we see that it is the (right) eigenvector of $\mathcal{%
G}$\ with unit eigenvalue. Meanwhile, the joint distribution in this state
will be homogeneous in time and so, depends only on the difference $\tau
\equiv T^{\prime }-T:$
\begin{equation*}
\mathcal{P}^{\ast }(\vec{n}^{\prime },T^{\prime };\vec{n},T)=\mathcal{G}%
^{\tau }(\vec{n}^{\prime },\vec{n})P^{\ast }(\vec{n})
\end{equation*}

\section{Microscopic Characterization of the NESS}

As summarized in Appendix~\ref{app:AppendixA}, a significant consequence of
detailed balance violation is that the system relaxes into a \textit{%
non-equilibrium steady state} (NESS). Typically, finding the associated
stationary distribution $P^{\ast }$ is a challenging task~\footnote{%
Though a systematic procedure, given $\mathcal{G}$,\ to find $P^{\ast }$
exists~\cite{hill1966studies}, it is too cumbersome to be practical in typical cases.}. 
Further, due to the absence of time reversal, this NESS is characterized
by non-vanishing probability currents~\cite{zia2007probability}. Though \textit{net}
currents may flow between any two configurations in general, here they exist
only between nearest neighbors, e.g., from $\vec{n}$ to $\vec{n}+\vec{e}_{i}$%
. Thus, we can denote it by a vector field $\vec{K}^{\ast }$, the components
being%
\begin{eqnarray}
\label{Kstar}
K_{1}^{\ast }(\vec{n}) &=&W_{1}^{+}(\vec{n})P^{\ast }(\vec{n})-W_{1}^{-}(%
\vec{n}+\vec{e}_{1})P^{\ast }(\vec{n}+\vec{e}_{1}),  \label{eqn:Kast} \\
K_{2}^{\ast }(\vec{n}) &=&W_{2}^{+}(\vec{n})P^{\ast }(\vec{n})-W_{2}^{-}(%
\vec{n}+\vec{e}_{2})P^{\ast }(\vec{n}+\vec{e}_{2}),  \notag
\end{eqnarray}%
Note that $-K_{i}^{\ast }(\vec{n}-\vec{e}_{i})$ is also the net current from 
$\vec{n}$ to $\vec{n}-\vec{e}_{i}$. In a NESS, the current is
divergence-free (see Appendix~\ref{app:AppendixA}). On a lattice, this
condition reads 
\begin{equation}
0=K_{1}^{\ast }(\vec{n})-K_{1}^{\ast }(\vec{n}-\vec{e}_{1})+K_{2}^{\ast }(%
\vec{n})-K_{2}^{\ast }(\vec{n}-\vec{e}_{2}).  \label{divK}
\end{equation}%
As a result, the curl of $\vec{K}^{\ast }$ is non-trivial, i.e., $\vec{K}
^{\ast }$ forms closed loops, a useful characterization of which is the
vorticity, $\omega ^{\ast }$ (see Appendix~\ref{app:AppendixA}). On the discrete $S_{1}\times S_{2}$ lattice,
any closed loop can be regarded as the sum of elementary loops, each around
a square (plaquette). Thus, we associate $\omega ^{\ast }$ with a plaquette
instead of a site. While the center of a plaquette is located at half
integers $(n_{1}+1/2,n_{2}+1/2)$~\cite{mellor2016supp}, we will still use $\vec{n}$
for convenience. Thus we write%
\begin{equation}
\omega ^{\ast }\left( \vec{n}\right) =K_{1}^{\ast }(\vec{n})+K_{2}^{\ast }(%
\vec{n}+\vec{e}_{1})-K_{1}^{\ast }(\vec{n}+\vec{e}_{2})-K_{2}^{\ast }(\vec{n}%
).  \label{eqn:omega}
\end{equation}%
Of course, $\omega ^{\ast }$ is related to the (generalized) discrete
Laplacian of $P^{\ast }$ and so, carries information about the curvature of $%
P^{\ast }$. At the intuitive level, such loops also suggest that one species
\textquotedblleft following\textquotedblright , or \textquotedblleft
chasing\textquotedblright , the other (see Fig.~\ref{fig:steady_state}%
(b,c)), much like the time-irreversible dynamics of predator-prey systems.

Apart from the vorticity, it is useful to characterize a divergence free
vector field as the curl of another. Since our space is two dimensional,
this field is a scalar, $\psi ^{\ast }$, known in fluid dynamics as the
stream function. Thus, 
\begin{equation}
K_{i}^{\ast }(\vec{n})=\varepsilon _{ij}\left[ \psi ^{\ast }\left( \vec{n}%
\right) -\psi ^{\ast }\left( \vec{n}-\vec{e}_{j}\right) \right]
\label{eqn:K-psi}
\end{equation}%
where $\varepsilon _{ij}$ is the two-dimensional Levi-Civita symbol and
repeated indices are summed. On our lattice, $\psi ^{\ast }$ is also
associated with a plaquette so that we can use the same scheme as in $\omega
^{\ast }$. If we chose the arbitrary constant in $\psi ^{\ast }$ to be zero
just outside the $S_{1}\times S_{2}$ rectangle, then we find%
\begin{equation}
\psi ^{\ast }\left( \vec{n}\right) =\sum_{\ell =0}^{n_{2}}K_{1}^{\ast
}(n_{1},\ell )=-\sum_{\ell =0}^{n_{1}}K_{2}^{\ast }(\ell ,n_{2}),
\label{eqn:stream}
\end{equation}%
For the more familiar continuum versions of $\vec{K}^{\ast }$, $\omega
^{\ast }$, and $\psi ^{\ast }$; see Appendix~\ref{app:AppendixA2}.

Finally, in formal analogy with fluid dynamics, we can associate these
probability current loops with the concept of \emph{probability angular
momentum.} As $\vec{L}=\int_{\vec{r}}\vec{r}\times \vec{J}\left( \vec{r}%
\right) $ represents the total angular momentum of a fluid with current
density $\vec{J}$, the sum $\sum_{\vec{n}}\left[ n_{1}K_{2}^{\ast }(\vec{n}%
)-n_{2}K_{1}^{\ast }(\vec{n})\right] $ plays the same role in the 2$q$VZ
case. Since $\vec{K}^{\ast }\propto P^{\ast }$, such a sum can be recognized
as a form of statistical average. It is reasonable to label it as the 
\textit{average} total probability angular momentum (in the NESS):%
\begin{equation}
\langle \mathcal{L}\rangle ^{\ast }\equiv \sum_{\vec{n}}\varepsilon
_{ij}n_{i}K_{j}^{\ast }(\vec{n}).  \label{eqn:L2qVZ}
\end{equation}%
In this context, it is possible to regard $\vec{K}^{\ast }$ as a kind of
probability distribution, much like $P^{\ast }$. In fact, substituting (\ref%
{eqn:Kast}) into this expression leads us to 
\begin{equation}
\langle \mathcal{L}\rangle ^{\ast }=\left\langle \varepsilon
_{ij}n_{i}V_{j}\right\rangle ^{\ast }  \label{eqn:LnV}
\end{equation}%
where $\vec{V}\equiv \vec{W}^{+}-\vec{W}^{-}$; see Appendix~\ref{app:AppendixA2}. Clearly, $\langle \mathcal{L}%
\rangle ^{\ast }$ is as much as a physical observable in the NESS as the
mean $\langle n_{i}\rangle ^{\ast }\equiv \sum_{\vec{n}}{n_{i}P^{\ast }(\vec{%
n})}$ or the two-point lagged correlations $\langle n_{i}^{\prime
}n_{j}\rangle _{T}^{\ast }\equiv \sum_{\vec{n},\vec{n}^{\prime }}{n}^{\prime
}{_{i}n_{j}\mathcal{P}^{\ast }(\vec{n}^{\prime },T;\vec{n},0)}$. Indeed, in
the limit of large $N$, we will see that $\langle \mathcal{L}\rangle ^{\ast
} $ is directly related to the \emph{antisymmetric} part of $\langle
n_{i}^{\prime }n_{j}\rangle _{T=1}^{\ast }$~\cite{shkarayev2014exact,mellor2016characterization}, see (\ref
{PAM}). In subsequent sections, we will devote much of our attention to such
quantities.

\section{Numerically Exact Results for Small Systems}

We have already noted that for systems in NESS, it is difficult to find the exact analytic expressions for $
P^{\ast }$ in general. However, for small systems, it is possible to attain numerically the 
\textquotedblleft exact\textquotedblright\  $
P^{\ast }$ to a
prescribed accuracy. Once $P^{\ast }(\vec{n})$ is known, with (\ref{Kstar})-(\ref{eqn:L2qVZ}), it is then
straightforward to compute all other quantities of interest, e.g., the
stationary probability current $K^{\ast }(\vec{n})$, vorticity $\omega ^{\ast
}(\vec{n})$, stream function $\psi ^{\ast }(\vec{n})$, angular momentum $%
\langle \mathcal{L}\rangle ^{\ast }$, and any correlation function.
It should be clear that in this subsection the evolution operator $\mathcal{G}$ is an $(S_1+1)(S_2+1)\times (S_1+1)(S_2+1)$ matrix and we consider here the evolution according to (\ref{eqn:ME}) in matrix form, i.e.,  $\vec{P} (T +1) =\mathcal{G} \vec{P} (T)$, where $\vec{P} (T)$ is the probability vector  whose elements are $P (\vec{n}, T )$.

To find a numerically exact $P^{\ast }(\vec{n})$ from the evolution operator
(\ref{Gexpl}), we thus exploit the matrix relation (\ref{eqn:ME}) $\mathcal{G}^{\infty }\vec{P}(0)=\vec{P}(\infty)=\vec{P}^*$
independently of $\vec{P}(0)$.
In practice, we compute $\vec{P}^{\ast }=\mathcal{G}^{\infty }\vec{P}(0)$
 by iterating  $\mathcal{G}^{2\tau }=\mathcal{G}^{\tau }\mathcal{G}^{\tau }$ until the desired accuracy is reached. In particular, for a system
with $S_{1}=S_{2}=100$, we find $P^{\ast }$ accurate up to $10^{-15}$ with
just $64$ iterations (i.e., $\mathcal{G}^{2^{64}}$). Since there are $
(S_{1}+1)\times (S_{2}+1)$ possible states, the total number of entries in
the matrix $\mathcal{G}$ is $[(S_{1}+1)(S_{2}+1)]^{2}$, which is $O\left(
10^{8}\right) $ for $S_{1}=S_{2}=100$. Of course, $\mathcal{G}$ is sparse,
as there are only five transitions from each state. However subsequent
powers of $\mathcal{G}$ soon become dense. This is problematic for
calculating the stationary distribution for large systems where one requires
a trade off between storing higher powers of $\mathcal{G}$ and the
computational cost of repeated multiplications of $\vec{P}$ by $\mathcal{G}$ to
some power, in order to reach the stationary state quickly. Nevertheless,
for systems larger than $S=100$, it is possible to attain accurate $\vec{P}^{\ast
} $'s relatively quickly by exploiting a method which combines iteration and
interpolation (see Appendix \ref{app:numerics}).

\begin{figure}[tbp]
\centering
\includegraphics[width=\linewidth]{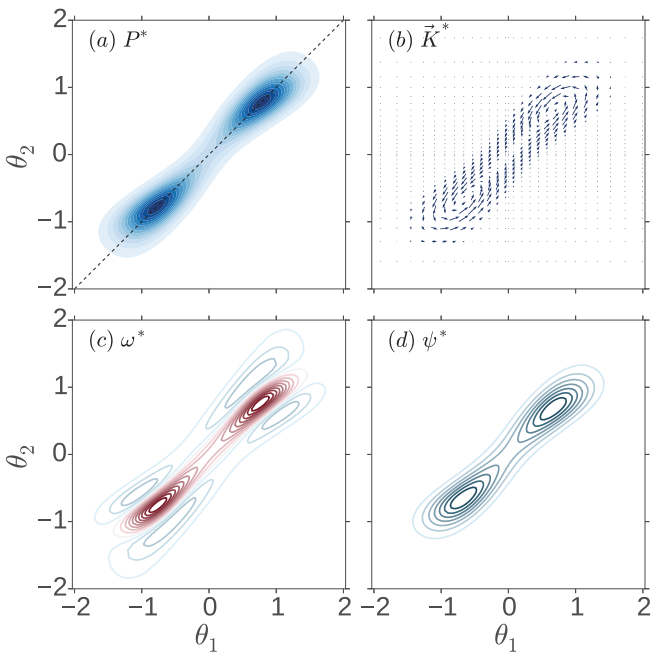}
\caption{\textit{(Color online)}. Exact properties of the 2$q$VZ NESS with $%
(N,S,Z,q_1,q_2)=(280,100,40,1,2)$ (low-zealotry phase). (a) Heat maps
of the stationary distribution $P^{\ast}$ in $\protect\theta$-space where $%
P^*$'s peaks are on the dotted line $\protect\theta_1=\protect\theta_2$, see
text. Areas of higher probability appear darker. (b) Stationary
probability currents, $\vec{K}^{\ast}$. (c) Stationary vorticity $%
\protect\omega^{\ast}$ in $\protect\theta$-space. Regions where $\protect%
\omega^{\ast}$ is positive/negative are in red/blue (dark/light in
grayscale). (d) Stationary stream function, $\protect\psi^{\ast}$%
.}
\label{fig:steady_state}
\end{figure}

As briefly explained in Ref.~\cite{mellor2016characterization}, the stationary distribution of
the 2$q$VZ is characterized by two phases in general (as in the $q$VMZ~\cite{mobilia2015nonlinear}): When the fraction of zealots is low, the long-time opinion
distribution $P^{\ast }$ is bimodal (\textit{low-zealotry phase}) whereas it
is single-peaked when the density of zealotry exceeds a critical value (%
\textit{high-zealotry phase}). While this scenario will be scrutinized in
the next sections, it is useful to gain some insight by considering some
typical systems.

In many ways, the qualitative behavior of the 2$q$VZ resembles that of the
mean-field version of the two-dimensional Ising model. The sum and
difference, $\left( Z_{+}+Z_{-}\right) /N$ and $\left( Z_{+}-Z_{-}\right) /N$%
, play the role of temperature and external magnetic field, respectively.
Thus, the transition in $P^{\ast }$ from displaying a single peak to being
bimodal as zealotry is lowered. In the latter case, the two peaks are of
equal height only when $Z_{+}=Z_{-}$ and the system exhibits \textit{
criticality}. Similar to the Ising model in a magnetic field, the 2$q$VZ is
no longer critical when $Z_{+}\neq Z_{-}$ and the bimodal stationary
probability is characterized by one peak that vanishes with the system size.
While this phenomenology has also been observed in the $q$VMZ~\cite{mobilia2015nonlinear},
the 2$q$VZ is genuinely out-of-equilibrium and it settles into a NESS whose
quantitative features are shaped by non-trivial probability currents.

Using the methodology outlined above, we have explored the NESS of systems
of various sizes up to $S_{1}\times S_{2}=100\times 100$. For the sake of illustration, we
show typical results for $P^{\ast }$, $\vec{K}^{\ast }$, $\omega ^{\ast }$,
and $\psi ^{\ast }$ in Fig.~\ref{fig:steady_state}, for the \textit{low
zealotry} phase in the symmetric case $S_{1}=S_2=100$ and $Z_{\pm }=40$. Of
course, the natural variables are just $n_{i}/S_{i}$, the fractions of each
population holding opinion $+1$. On the other hand, there are some
advantages to using a different set of variables: $\theta _{i}=\left(
1/q_{i}\right) \ln {\left[(S_{i}/n_{i})-1\right]}$. First, the physical $\theta $'s
occupy the entire plane, i.e., $\theta\in (-\infty,\infty)$, while the Ising-like symmetry of the 2$q$VZ
corresponds to $\theta _{i}\leftrightarrow -\theta _{i}$. Second, the
symmetry point $n_{i}=S_{i}/2$, is mapped to the origin. Finally, though
there is no symmetry under the exchange of the two populations ($
i=1\leftrightarrow 2$), we see that $P^{\ast }(\vec{\theta})$ is nearly
symmetric under this reflection, see Fig.~\ref{fig:steady_state}(a). In
particular, as will be shown below, the fixed points of the deterministic
mean-field rate equations (\ref{eqn:rate_equation}), which correspond to the
peaks of $P^{\ast }$, are situated on the $\theta _{1}=\theta _{2}$ line.
Indeed, the \textquotedblleft ridge\textquotedblright\ which runs from one
peak through the saddle to the other is seen to lie very close to this line.
In Fig.~\ref{fig:steady_state}(b), we show the current field $\vec{K}^{\ast
} $, which whirls around each peak of $P^{\ast }$. These whirls imply
correlations of the dynamic properties of the two populations, indicating
the tendency of $q_{2}$-voters to \textquotedblleft
follow\textquotedblright\ or \textquotedblleft chase\textquotedblright\ $
q_{1}$-voters~\cite{mellor2016characterization}, a result that is corroborated by two-point
correlation functions in Sec.~IV.B. The bottom panels of Fig.~\ref
{fig:steady_state} show the associated vorticity field and stream function.
As expected, $\omega ^{\ast }$ is positive near the peaks, corresponding to
the current whirling counter-clockwise. Counter rotations ($\omega ^{\ast
}<0 $) are present away from the peaks, a property not readily discerned
when we examine the currents in Fig.~\ref{fig:steady_state}(b). Lastly, we
observe that the stream function $\psi ^{\ast }$ appears to be very similar
to $P^{\ast }$. This behavior is not a coincidence, as we will show that, in
the linear Gaussian approximation, the two are strictly proportional to each
other.

\section{Continuum Descriptions and Simulation Studies of Large Systems}

\label{sec:descriptions}

As is often the case, the exact description of the 2$q$VZ through the master
equation is intractable and difficult to analyze when the system size is
large. Typically, Monte Carlo simulations are needed to explore their
behavior. However, much insight can be gained through a continuum
description and judicious approximations. Specifically, we consider the
thermodynamic limit: $Z_{\pm },S_{i},N\rightarrow \infty $, with fixed
densities $z_{\pm }=Z_{\pm }/N$ and $s_{i}=S_{i}/N$. In this limit, we
expect demographic fluctuations to be negligible and the description in
terms of mean-field rate equations is suitable. For large but finite
systems, we can account for  fluctuations and correlations via the Fokker-Planck equation
(FPE)~\cite{vankampen1992stochastic,*gardiner1985handbook,*risken1984fokker} associated with Eq.~(\ref{eqn:ME}).

\subsection{The Fokker-Planck Equation}

When $N\gg 1$, the configuration space of densities $x_{i}=n_{i}/N$
approaches the continuum within a rectangle: $x_{i}\in \left[ 0,s_{i}\right] 
$, while time is rescaled so that $t=T/N$ is continuous. In other words, we
will say that one Monte Carlo step (1 MCS) corresponds to $N$ moves of the
microscopic model. Following standard procedures~\cite{lax1966classical} to
obtain the continuum limit of the ME (\ref{eqn:ME}), we arrive at the FPE
for the probability density $P\left( \vec{x};t\right) $: 
\begin{equation}
\frac{\partial }{\partial t}P\left( \vec{x};t\right) =\sum_{i=1,2}\frac{%
\partial }{\partial x_{i}}\left[ \frac{\partial }{\partial x_{i}}u_{i}(\vec{x%
})P\left( \vec{x};t\right) -v_{i}(\vec{x})P\left( \vec{x};t\right) \right]
\label{eqn:FPE}
\end{equation}%
where $u_{i}\equiv \left( w_{i}^{+}+w_{i}^{-}\right) /2N$ and $v_{i}\equiv
w_{i}^{+}-w_{i}^{-}$, and where $%
w_{i}^{+}=(s_{i}-x_{i})(z_{+}+x_{1}+x_{2})^{q_{i}}$ and $%
w_{i}^{-}=x_{i}(z_{+}+x_{1}+x_{2})^{q_{i}}$ are the continuum counterparts
of $W_{i}^{\pm }$. 
The right-hand-side (RHS) can be identified as the divergence of the \emph{
probability current density} 
\begin{equation}
K_{i}(\vec{x};t)=v_{i}P\left( \vec{x};t\right) -\partial _{i}\left[
u_{i}P\left( \vec{x};t\right) \right]  \label{eqn:K-FPE}
\end{equation}%
where $\partial _{i}\equiv \partial /\partial x_{i}$. Clearly, in the NESS $%
P\left( \vec{x};t\right) \rightarrow P^{\ast }\left( \vec{x}\right) $,
whereas the corresponding stationary probability current density is $%
K_{i}^{\ast }(\vec{x})=v_{i}P^{\ast }\left( \vec{x}\right) -\partial _{i}%
\left[ u_{i}P^{\ast }\left( \vec{x}\right) \right] $. To find $P^{\ast }$,
we must solve the partial differential equation $\sum_{i=1,2}\partial _{i}%
\left[ \partial _{i}u_{i}(\vec{x})P^{\ast }\left( \vec{x}\right) -v_{i}(\vec{%
x})P^{\ast }\left( \vec{x}\right) \right] =0$ with non-trivial boundary
conditions: vanishing of the normal components of $\vec{K}^{\ast }$. Thus,
obtaining $P^{\ast }$ analytically in general is still quite challenging.

\subsection{Mean-Field Analysis}

A standard alternative to studying the FPE for the full distribution is to
consider the equations governing the evolution of averages of various
quantities~\cite{bogolubov1947kinetic,*yvon1935theorie}. In particular, it is intuitively interesting to
study the behavior of the average $\left\langle x_{i}\right\rangle
_{t}\equiv \int x_{i}P\left( \vec{x};t\right) d\vec{x}$. Its evolution is
governed by%
\begin{equation*}
\frac{d}{dt}\left\langle x_{i}\right\rangle =\int x_{i}\frac{\partial }{%
\partial t}P\left( \vec{x};t\right) d\vec{x}=\int K_{i}\left( \vec{x}; t\right)
d\vec{x}
\end{equation*}%
since the surface contributions from the integration by parts involve the
normal components of $\vec{K}$ and vanish. To make progress, we will
need to make approximations. First, $\int \partial _{i}\left[ u_{i}P\right] $
is also a surface term. Though it is not necessarily zero, it vanishes at
the lowest order for large $N$. In this limit, we are left with contributions arising
only from the
first term on the right-hand-side of (\ref{eqn:K-FPE}): 
\begin{eqnarray}
&&\frac{d}{dt}\left\langle x_{i}\right\rangle =\left\langle v_{i}\left( \vec{%
x}\right) \right\rangle  \label{RE0} \\
&=&\langle (s_{i}-x_{i})(z_{+}+x_{1}+x_{2})^{q_{i}}\rangle -\langle
x_{i}(z_{+}+x_{1}+x_{2})^{q_{i}}\rangle .  \notag
\end{eqnarray}%
Second, we invoke the mean-field approximation (MFA), which assumes that
higher moments can be factored in terms of the averages $\langle
x_{i}\rangle $ (e.g., $\left\langle x_{i}x_{j}\right\rangle \approx
\left\langle x_{i}\right\rangle \left\langle x_{j}\right\rangle $). In this
sense, the MFA neglects all correlations and fluctuations, an approach that becomes exact
when $N\rightarrow \infty $. Within this approach, Eq.~(\ref{RE0}) becomes
the mean-field rate equations (REs) 
~\footnote{For notational convenience, the variables in (\ref{eqn:rate_equation}) are the averages \unexpanded{$\left\langle \vec{x}\right\rangle$}. There should be no confusion with components of $\vec{x}$ in the configuration space.}
: 
\begin{equation}
\frac{d}{dt}x_{i}=(s_{i}-x_{i})\mu ^{q_{i}}-x_{i}(1-\mu )^{q_{i}}
\label{eqn:rate_equation}
\end{equation}%
where $\mu =z_{+}+x_{1}+x_{2}$ is the density of voters holding opinion $+$.

The fixed points (FPs) of these rate equations are stable or unstable nodes
(no limit cycles in our case~\cite{mellor2016characterization,mellor2016supp}). They are given by 
\begin{equation}
x_{i}^{\ast }=\frac{s_{i}}{1+\rho ^{q_{i}}},  \label{eqn:x*}
\end{equation}%
where $\rho =\left( 1-\mu ^{\ast }\right) /\mu ^{\ast }$, with $\mu ^{\ast }=z_{+}+x_{1}^{\ast }+x_{2}^{\ast }$. 
Note that $\rho $
is the ratio, in the steady state, of voters holding opinion $-$ to those
with the $+$ opinion. It satisfies 
\begin{equation}
\frac{1}{1+\rho }=z_{+}+\sum_{i=1,2}\frac{s_{i}}{1+\rho ^{q_{i}}}
\label{eqn:rho}
\end{equation}%
since the left hand side is $\mu ^{\ast }=z_{+}+x_{1}^{\ast }+x_{2}^{\ast }$%
, which equals the right hand side. In terms of the variables $\theta _{i}$
introduced above, we recognize that $x_{i}=s_{i}/\left( 1+e^{q_{i}\theta
_{i}}\right) $, so that all FPs are given by $\theta _{i}^{\ast }=\ln \rho $%
.

We are particularly interested in the case $z_{+}=z_{-}=z$ for which the 2$q$VZ exhibits a continuous phase transition. 
In this case, it is clear that $
\rho =1$ ($\mu ^{\ast }=1/2$) is always a solution to Eq.~(\ref{eqn:rho}),
corresponding to the \textquotedblleft central FP\textquotedblright : $\vec{x
}^{\ast }=\vec{x}^{(0)}\equiv (s_{1}/2,s_{2}/2)$.The properties of this FP
changes, as $z$ is decreased below a critical value $z_{c}$, from being
stable to unstable. To show this property, we linearize Eq.~(\ref
{eqn:rate_equation}) about a FP $\vec{x}^{\ast }$ and find the linear
stability matrix, $-(\partial \dot{x_{i}}/\partial x_{j})|_{\vec{x}=\vec{x}%
^{\ast }}$, to have the form 
\begin{equation}
\mathbb{F}(\vec{x}^{\ast })=\left( 
\begin{array}{cc}
Y_{1\mu }-X_{1\mu } & -X_{1\mu } \\ 
-X_{2\mu } & Y_{2\mu }-X_{2\mu }%
\end{array}%
\right) ,  \label{eqn:F}
\end{equation}%
where 
\begin{eqnarray*}
Y_{i\mu } &=&\mu ^{\ast {q_{i}}}(1+\rho ^{q_{i}}), \\
X_{i\mu } &=&{q_{i}}x_{i}^{\ast }(1-\mu ^{\ast })^{{q_{i}-1}}\left( 1+\rho
\right) .
\end{eqnarray*}%
Evaluating $\det {\mathbb{F}}$ at $\vec{x}^{\ast }={x^{(0)}}$, we find a
remarkably simple result:%
\begin{equation}
\det {\mathbb{F}(\vec{x}^{(0)})}=2^{2-q_{1}-q_{2}}\left[ 1-{q_{2}}s_{2}-{%
q_{1}}s_{1}\right]  \label{detF0}
\end{equation}%
Since a stable FP is associated with $\det {\mathbb{F}(\vec{x}^{(0)})}>0$,
this expression implies that the 2$q$VZ resembles an Ising model at high
temperatures when 
\begin{equation}
1>s_{1}q_{1}+s_{2}q_{2}~~.  \label{Stability}
\end{equation}%
In the case $\det {\mathbb{F}(\vec{x}^{(0)})}<0$, ${\vec{x}^{(0)}}$ turns
unstable and we can verify that there are two other (real $\rho $) solutions to
Eq.~(\ref{eqn:rho}), corresponding to two other\ stable, \textquotedblleft
non-central\textquotedblright\ FPs: $\vec{x}^{(\pm )}$ Of course, these
correspond to the low temperature phase of the Ising model, with a
spontaneously broken symmetry. Thus, we will refer to the line $%
s_{1}q_{1}+s_{2}q_{2}=1$ as the critical line (plotted in Fig.~\ref%
{fig:critical_z}), separating the $s_{1}$-$s_{2}$ plane into a region where $\vec{
x}^{(0)}$ is the sole FP and another where Eq.~(\ref{eqn:rate_equation})
admits three FPs. Before continuing, we emphasize that this technique is
powerful enough for us to obtain the generalization of (\ref{Stability}) to
a population with any number of groups of $q_{i}$-voters, namely, $\vec{
x}^{(0)}$ is  stable when $1>\Sigma
_{i}s_{i}q_{i}$.
Details are given  in Appendix \ref{app:z_critcal}.

Since $1=2z+s_{1}+s_{2}$, we can express the critical $z_{c}$ as a function
of the $q$'s and the difference $\Delta s\equiv s_{1}-s_{2}$ 
\begin{equation}
z_{c}=\frac{\bar{q}-1}{2\bar{q}}+\frac{q_{1}-q_{2}}{q_{1}+q_{2}}\frac{\Delta
s}{2},  \label{zc}
\end{equation}
where $\bar{q}$ is the average $\left( q_{1}+q_{2}\right) /2$. In the limit
of a homogeneous population ($q_{1}=q_{2}$), we recover the critical values $
z_{c}=(q-1)/(2q)$ of the $q$VMZ~\cite{mobilia2015nonlinear}. In this spirit, we may
introduce an \textit{effective} $q_{\text{eff}}$ for our heterogeneous $2q$VZ:
\begin{equation}
q_{\text{eff}}=\frac{s_{1}q_{1}+s_{2}q_{2}}{s_{1}+s_{2}}  \label{eqn:q_eff}
\end{equation}%
in the sense that the number of FPs in the $2q$VZ are the same as those in
the homogeneous $q$VMZ with this $q_{\text{eff}}$. Indeed, as shown in
Appendix \ref{app:z_critcal}, this notion can be generalized to $q_{\text{eff%
}}=\left. \left( \Sigma _{i}s_{i}q_{i}\right) \right/ \left( \Sigma
_{i}s_{i}\right) $ in the case of populations with any composition. 
\begin{figure}[tbp]
\centering
\includegraphics[width=0.6\linewidth]{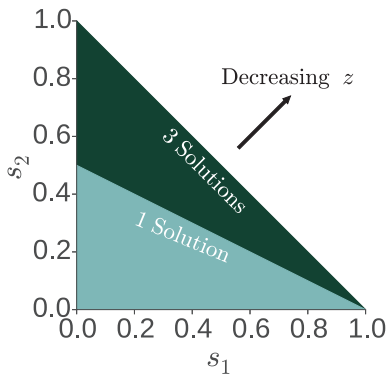}
\caption{\textit{(Color online).} Criticality in the $2q$VZ ($z_{\pm }=z$):
Dependence of $z_{c}$ on the susceptible densities. 
Criticality at $z_{c}$ occurs on the
interface between the blue/grey and green/black regions, prescribed by $
s_{1}q_{1}+s_{2}q_{2}=1$, and the corresponding critical zealotry density is 
$z_{c}=(1-s_{1}-s_{2})/2$; see text. In the blue/grey region, the mean-field equations (\ref{eqn:rate_equation})
have one fixed point, whereas they admit three fixed points in the green/black region; see text.
Here, $(q_{1},q_{2})=(1,2)$.}
\label{fig:critical_z}
\end{figure}

\subsection{Beyond MFA for Systems Far from Criticality}

\label{sec:fluctuations}

While the MFA provides an adequate picture of the deterministic aspects of
the system's evolution and the location of a phase transition, it cannot
capture the stochastic nature of our model. In this section, we study the
influence of demographic fluctuations in finite populations by means of the
linear Gaussian approximation (LGA) of the FPE (\ref{eqn:FPE})~\cite{zia2007probability,mellor2016characterization}. To be concise, we 
focus on systems with symmetric zealotry ($
z_{\pm }=z$) far from criticality, i.e., $z\ll z_{c}$ and $
z\gg z_{c}$. In the realm of the LGA, we are able to compute many quantities
of interest exactly, in the NESS. Beyond the LGA, we also study the mean
switching time that characterizes the stochastic dynamics in the low
zealotry phase when the system endlessly switches between the two stable
fixed points $\vec{x}^{(\pm )}$. %

\subsubsection{Linear Gaussian Approximation of $P^{\ast }$}

To describe fluctuations in the NESS, we examine the FPE (\ref{eqn:FPE})
beyond the lowest order in $1/N$. This procedure can provide a scaling
analysis of typical non-critical fluctuations in systems of large but finite
size $N$. As our focus is the behavior in a NESS, we consider small  deviations near a (stable)
fixed point $\vec{x}^{\ast }$: $\vec{\xi}=\vec{x}-\vec{
x}^{\ast }$. The LGA consists of linearizing the drift term of (\ref{eqn:FPE}), i.e., $v_{i}\rightarrow (-\mathbb{F}\cdot \vec{\xi})_{i}$, while the
diffusion coefficient is evaluated at $\vec{x}^{\ast }=(x_{1}^{\ast
},x_{2}^{\ast })$, i.e., $u_{i}\rightarrow w_{i}^{\ast }=w_{i}^{+}(\vec{x}%
^{\ast })=w_{i}^{-}(\vec{x}^{\ast })$. Upon substitution into FPE (\ref%
{eqn:FPE}), we obtain the LGA-FPE for the \ stationary distribution: 
\begin{equation}
0=\sum_{i,j=1,2} 
\frac{\partial }{\partial \xi _{i}} \left(
\frac{\partial }{\partial \xi _{j}}
D_{ij}+F_{ij}\xi _{j} \right)
P^{\ast }(\vec{\xi}).  \label{eqn:stationary_FPE} 
\end{equation}
Here, $D_{ij}$ and $F_{ij}$ are the elements of the diffusion matrix $%
\mathbb{D}(\vec{x}^{\ast })$ and the drift matrix $\mathbb{F}(\vec{x}^{\ast
})$, respectively. The former is given by%
\begin{equation}
\mathbb{D}(\vec{x}^{\ast })=\frac{1}{N}\left( 
\begin{array}{cc}
x_{1}^{\ast }(1-\mu ^{\ast })^{q_{1}} & 0 \\ 
0 & x_{2}^{\ast }(1-\mu ^{\ast })^{q_{2}}%
\end{array}%
\right)  \label{Ddiff}
\end{equation}%
(i.e., $D_{ij}=\delta _{ij}w_{i}^{\ast }/N$), while $\mathbb{F}$ is given by
(\ref{eqn:F})~\footnote{%
If $\mathbb{D}\mathbb{F}^{-1}$ were symmetric, then detailed balance would be
satisfied, which is not the case here.}. In general, the solution of (\ref%
{eqn:stationary_FPE}) is still a Gaussian~\cite{lax1966classical}: 
\begin{equation}
P^{\ast }(\vec{\xi})\propto \mathrm{exp}\left[ -\frac{1}{2}\vec{\xi}\cdot 
\mathbb{C}^{-1}\vec{\xi}\right] ,  \label{eqn:LGA2}
\end{equation}%
where $\mathbb{C}$ is the covariance matrix, with elements%
\begin{equation}
C_{ij}=\langle \xi _{i}\xi _{j}\rangle ^{\ast },  \label{Cij}
\end{equation}%
and can be obtained from $\mathbb{D}$ (\ref{Ddiff}) and $\mathbb{F}$ (\ref%
{eqn:F}) by solving $\mathbb{F}\mathbb{C}+\mathbb{C}\mathbb{F}^{T}=2\mathbb{D%
}$~\cite{weiss2003coordinate, *weiss2007fluctuation}. Details, as well as the explicit forms for $\mathbb{C}$ in
our case can be found in Appendix~\ref{app:explicit}. Since $\mathbb{D}$ is $%
O\left( 1/N\right) $ and $\mathbb{F}$ is $O\left( 1\right) $, it follows
that $\mathbb{C}$ is also $O\left( 1/N\right) $, confirming our expectation
that the fluctuations (standard deviations) are $O\left( 1/\sqrt{N}\right) $%
. Needless to say, if we approach criticality, then one of the eigenvalues
of $\mathbb{F}$ approaches zero, so that $\mathbb{C}$ diverges and this
approximation breaks down. On the other hand, for fixed $z\neq z_{c}$, the
accuracy of the Gaussian expression (\ref{eqn:LGA2}) improves as $%
N\rightarrow \infty $. As an example of the quality of the LGA, we find
that, for $N=400$ and $S_{i}=Z_{\pm }=100$, the prediction from (\ref%
{eqn:LGA2}) with (\ref{eqn:C0}) agrees with the numerically exact $P^{\ast }$
to $\lesssim $ $2\%$ within about two standard deviations.

\subsubsection{Currents and correlations in the LGA\protect\footnote{%
In this subsection, we use Einstein's summation convention for repeated
indices. }}

With a known $P^{\ast }$, we can find two exact expressions for the currents 
$\vec{K}^{\ast }=-\left\{ \mathbb{D}\vec{\nabla}+\mathbb{F}\vec{\xi}\right\}
P^{\ast }$. One displays the linear relationship $\vec{K}^{\ast }\propto 
\vec{\xi}P^{\ast }$:%
\begin{equation}
\vec{K}^{\ast }(\vec{\xi})=\left\{ \mathbb{D}\mathbb{C}^{-1}-\mathbb{F}%
\right\} \vec{\xi}P^{\ast }(\vec{\xi}).  \label{eqn:lga_currents}
\end{equation}%
The other shows explicitly that $\vec{K}^{\ast }$ is divergence free: 
\[
\vec{K}^{\ast }(\vec{\xi})=\frac{\mathbb{FC}-\mathbb{C}\mathbb{F}^{T}}{2}%
\vec{\nabla}P^{\ast }(\vec{\xi}),
\]%
where we have used $\mathbb{D=}\left[ \mathbb{F}\mathbb{C}+\mathbb{C}\mathbb{%
F}^{T}\right] /2$. The key observation here is that the matrix is
antisymmetric, which leads us to define%
\begin{equation}
\mathbb{L}=\left( 
\begin{array}{cc}
0 & L_{12} \\ 
-L_{12} & 0%
\end{array}%
\right) \equiv \mathbb{FC}-\mathbb{C}\mathbb{F}^{T}  \label{eqn:L}
\end{equation}%
In other words, $K_{i}^{\ast }=\varepsilon _{ij}\partial _{j}\left(
L_{12}P^{\ast }/2\right) $, allowing us to identify the stream function $%
\psi ^{\ast }=L_{12}P^{\ast }/2$, a linear relationship pointed out above.
Further, from the continuum version of (\ref{eqn:L2qVZ}), we see that~%
\footnote{%
We have ignored the surface terms in the integration by parts here, which
lies within the validity (and errors) of the Gaussian approximation.}%
\begin{equation}
\left\langle \mathcal{L}\right\rangle ^{\ast }=\int \varepsilon _{ij}\xi
_{i}K_{j}^{\ast }d\vec{\xi}=L_{12}  \label{eqn:L12}
\end{equation}%
Meanwhile, the vorticity field is proportional to $\xi _{i}\xi _{j}P^{\ast }$%
, so that it can be both positive and negative, as illustrated in Fig.~\ref%
{fig:steady_state}(c).

Finally, we turn our attention to the continuum version of the general two
point correlation, namely, $\langle \xi _{i}^{\prime }\xi _{j}\rangle _{\tau
}^{\ast }=\int \xi _{i}^{\prime }\xi _{j}\mathcal{P}^{\ast }\left( \vec{\xi}^{\prime
},\tau ;\vec{\xi},0\right) d\vec{\xi}^{\prime }d\vec{\xi}$. In the LGA, it
is much easier to use the solution to the corresponding Langevin equation \cite{vankampen1992stochastic,*gardiner1985handbook,*risken1984fokker}: $\vec{\xi}%
\left( \tau \right) =e^{-\mathbb{F}\tau }\vec{\xi}\left( 0\right) $ plus
noise. Since the noise is uncorrelated in time, 
\[
\left\langle \xi _{i}\left( 0\right) \xi _{j}\left( \tau \right)
\right\rangle ^{\ast }=e^{-F_{jk}\tau }\left\langle \xi _{i}\left( 0\right)
\xi _{k}\left( 0\right) \right\rangle ^{\ast }=C_{ik}e^{-F_{jk}\tau }
\]%
The antisymmetric part of this correlation is necessarily odd in $\tau $
and does not vanish for systems in NESS. Also central to our study of NESS,
we define the $t$-independent quantity~\cite%
{shkarayev2014exact,mellor2016characterization,zia2016manifest} 
\begin{equation}
\widetilde{C}(\tau )\equiv \langle \xi _{1}(0)\xi _{2}(\tau )\rangle
^{\ast }-\langle \xi _{2}(0)\xi _{1}(\tau )\rangle ^{\ast }
\end{equation}%
which is the $12$ element of the matrix%
\begin{equation}
\widetilde{\mathbb{C}}(\tau )=\mathbb{C}e^{-\mathbb{F}^{T}\tau }\mathbb{-}%
e^{-\mathbb{F}\tau }\mathbb{C}  \label{eqn:C-tilde}
\end{equation}
Since $v_{i}\left( 0\right) =\left. d\xi _{i}/d\tau \right\vert _{\tau =0}$,
we see that the continuum version of (\ref{eqn:LnV}) is $\left\langle 
\mathcal{L}\right\rangle ^{\ast }=\left\langle \varepsilon _{ij}\xi
_{i}v_{j}\right\rangle ^{\ast }$, which implies%
\begin{equation}
\left. \frac{d\widetilde{C}}{d\tau }\right\vert _{\tau =0}=\left\langle 
\mathcal{L}\right\rangle ^{\ast };~~\left. \frac{d\widetilde{\mathbb{C}}}{%
d\tau }\right\vert _{\tau =0}=\mathbb{FC}-\mathbb{C}\mathbb{F}^{T}
\label{C+L}
\end{equation}%
within the LGA. Thus, we can regard $\widetilde{C}(\tau )$ as a generalized
probability angular momentum, further details of which will be provided in
the next subsection.

\subsubsection{Simulation results for the symmetric case $s_{1,2}=s,z_{\pm
}=1/2-s$}

The above expressions (\ref{eqn:LGA2})-(\ref{eqn:C-tilde}) carry significant
information on the NESS of the 2$q$VZ in the realm of the LGA and are valid
for any values of $s_{1,2}$ and $q_{1,2}$. Here, we report the explicit results
obtained for the symmetric case $s_{1,2}=s,z_{\pm }=\frac{1}{2}-s$ ($%
q_{2}=2q_{1}=2$). These results complete those in Ref.~\cite{mellor2016characterization} and
allow us to discuss the validity of the LGA by considering a concrete
example. Here, the explicit expressions of $C_{ij}$, the stationary
correlation functions $\langle \xi _{i}\xi _{j}\rangle ^{\ast }$, are given
by (\ref{eqn:C0}) and (\ref{eqn:Cpm}), while the corresponding drift
matrices $\mathbb{F}$ have been given in Ref.~\cite{mellor2016supp} along with their
(real and distinct) eigenvalues labeled by $\lambda _{\pm }$, where $%
\lambda _{+}>\lambda _{-}$.

To assess the validity of the LGA, we have performed Monte Carlo simulations
of large systems with $N$ up to $14,400$, and compiled histograms from the
trajectories to find the probability distribution $P^{\ast }(\vec{n})$.
Similar to the comparison with the LGA prediction for smaller systems above,
we find that (\ref{eqn:LGA2}) is an excellent approximation in the
high-zealotry phase, i.e., $P^*(\vec{\xi})\simeq N P^*(\vec{n})$, with a sharp peak around the symmetric FP $\vec{x}
^{(0)} $. In the low zealotry phase, we have confirmed that the stationary
probability density is bimodal, with two sharp peaks close to the mean-field
FPs $\vec{x}^{(\pm )}$. However, the distribution around each FP~\footnote{
To compare with LGA predictions in the low zealotry phase, we ensure that
the trajectories never leave the vicinity of one of the two FPs. Naturally,
these runs complement those we use for estimating switching times.} is both
skewed and more sharply peaked than the LGA prediction. For systems with $
N\lesssim 10^{3}$ visible deviations between (\ref{eqn:LGA2}) and simulation
results exist. Yet, for larger systems ($N\gg 10^{3}$), the agreements are
quite reasonable, even in the low zealotry phase. We have confirmed this
analysis by computing the skewness and (excess) kurtosis of in the 1D
projections~\footnote{
For example, the projection onto the $\xi _{1}$-axis is given by $P^{*}(\xi
_{1})=\int_0^1 P^{*}(\xi _{1},\xi _{2}) d\xi_2$} of $P^{\ast }({\vec{\xi}})$ onto
each axis in both regimes. For the smallest system we considered ($S=250$),
in the low zealotry regime $Z=100$ the kurtosis was $(-0.242,0.750)$ for the 
$\xi _{1}$ and $\xi _{2}$ projections respectively. The skewness was $%
(0.321,0.969)$. For the high zealotry regime $Z=200$ the kurtosis was $%
(-0.030,-0.130)$ and skewness $(0.004,-0.007)$, confirming that the LGA is a
better approximation at the high zealotry regime.

\begin{figure}[tbp]
\centering
\includegraphics[width=\linewidth]{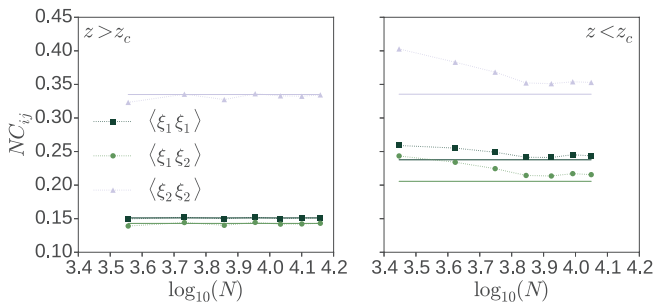} 
\caption{\textit{(Color online)}. $C_{ij}=\langle \protect\xi _{i}\protect%
\xi _{j}\rangle $ a function of system size $N$: Comparison of the LGA
predictions (\protect\ref{eqn:C0}) (left panel) and (\protect\ref{eqn:Cpm})
(right panel) in solid against results of stochastic simulations (markers), averaged
over at least $10^{5}$ MCS. The scaling $C_{ij}\propto 1/N$ is confirmed and
the quantitative agreement improves as $N$ increases (with reasonably good agreement in the
low-zealotry phase when $N\gg 10^{3}$).  Here, $s_{1,2}=s$, $z_{\pm }=z=\frac{1}{2}-s$
with $q_{2}=2q_{1}=2$. In the two regimes $z>z_{c}$ and $z<z_{c}$ we have $%
z=2/9$ and $z=1/7$ respectively, the critical value being $z_{c}=1/6$.}
\label{fig:correlations}
\end{figure}

\begin{figure}[tbp]
\centering
\includegraphics[width=\linewidth]{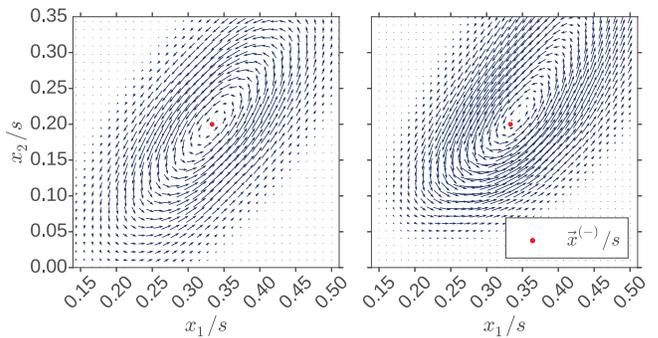}
\caption{\textit{(Color online)}. Stationary probability current: 
Comparison of the LGA predictions Eq.~(\protect\ref{eqn:lga_currents}) (left panel) with the numerically exact 
counterpart Eq.~(\ref{eqn:Kast}) (right panel) near the fixed point $x^{(-)}$ (red/grey dot) in the low
zealotry phase; see text. Parameters here are $%
(N,S_1, S_2,Z,q_{1},q_{2})=(280,100,100,40,1,2)$.}
\label{fig:K-omega}
\end{figure}

As the system size increases the kurtosis and skewness approach zero for
both regimes. In Fig.~\ref{fig:correlations}, the LGA predictions (\ref{eqn:C0})
and (\ref{eqn:Cpm}) in the high/low zealotry phases are compared against the
simulation results for $\langle \xi _{i}\xi _{j}\rangle ^{\ast }$. The
outcome confirms that $C_{ij}\propto 1/N$, i.e., fluctuations scale as $%
N^{-1/2}$, in both phases far from criticality. However, while the LGA
provides a good quantitative predictions for $N\gtrsim 10^{3}$ in the $
z>z_{c}$ cases, much larger system sizes ($N\gg 10^{3}$) are necessary to
reach a similar quantitative agreement in the low-zealotry phase. We believe
that the significant skewness associated with $\vec{x}^{(\pm )}$ in the
latter regime when $N$ is not sufficiently large is responsible for the differences. A better and
systematic understanding of this phenomenon is desirable, but beyond the
scope of this study. Using Eq.~(\ref{eqn:Kast}), we can compute the probability 
current $\vec{K}^{\ast }$ exactly (for small systems) or via simulations 
and compare the results with the current obtained from Eq.~(\ref{eqn:lga_currents}) by using $P^*$ obtained within the realm LGA as shown in Fig.~\ref{fig:K-omega}.
From this comparison, we notice that due to finite size effects 
the MF fixed point and peaks of the distribution (around which $\vec{K}^{\ast }$ whirls) do not coincide perfectly,
and the LGA flows are not symmetric around the fixed point. These discrepancies between simulations of the original 2$q$VZ and the LGA predictions
are attenuated and eventually
dissipate when the system size is increased.
Furthermore, we also verified that, as predicted by the LGA,  
the stream function is always positive, in agreement with the
counter-clockwise whirls of the probability current near the peaks reported
in Fig.~\ref{fig:steady_state}(b,c) and Fig.~\ref{fig:K-omega}. 

\begin{figure}[tbp]
\centering\includegraphics[width=\linewidth]{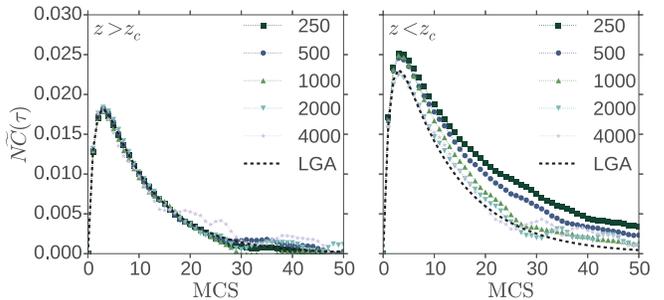}
\caption{\textit{(Color online)}. LGA prediction of the two point
correlation function, $\protect\widetilde{C}(\tau)$ (dashed) compared to
simulations (markers) in the high (left) and low (right) zealotry regimes for
the symmetric case $s_{1,2}=s$, $z_{\pm }=z=\frac{1}{2}-s$ with $q_{1}=1$
and $q_{2}=2$. In the high zealotry regime, the LGA captures the behavior
of $\protect\widetilde{C}$ for all values of $S$ considered. For the low
zealotry regime the LGA is qualitatively accurate for all $S$, however only
quantitatively accurate for $S\geq 2000$. Fluctuations for MCS greater than $%
20$ are due to a low sampling rate. The LGA values for the peak $\tau^{\ast }$
for the high and low zealotry regimes, $2.80$ an $3.05$ respectively,
accurately predict the data peak which occurs at $3$ in both cases.}
\label{fig:LGA}
\end{figure}

Turning to the antisymmetric two-point correlation function in the NESS $\widetilde{
\mathbb{C}}(\tau )$, we find $\widetilde{C}(\tau )$ from the simulation
trajectories $\vec{x}_{i}(t)$ via the lagged correlation: $\frac{1}{R-\tau }%
\sum_{t=0}^{R-\tau }\left[ x_{1}(t)x_{2}(t+\tau )-x_{2}(t)x_{1}(t+\tau )%
\right] $, where $R$ is the length of our run
(typically, $10^5$ MCS). From Eq.~(\ref{eqn:C-tilde}) we have%
\begin{equation*}
\widetilde{C}(\tau )=\langle \mathcal{L}\rangle ^{\ast }\left( \frac{%
e^{-\lambda _{-}\tau }-e^{-\lambda _{+}\tau }}{\lambda _{+}-\lambda _{-}}%
\right) 
\end{equation*}%
where $\lambda _{\pm }$ are the eigenvalues of $\mathbb{F}$. In Fig.~\ref%
{fig:LGA} we plot the LGA expression of $\widetilde{C}$ and the same
quantity obtained from stochastic simulations and again find a good
agreement in both high- and low-zealotry regimes. In particular, we notice
that the LGA accurately captures the peak of $\widetilde{C}(\tau)$ at $\tau
^{\ast }=[\ln {(\lambda _{+}/\lambda _{-})}]/[\lambda _{+}-\lambda _{-}]$~%
\cite{mellor2016characterization}. As illustrated in Fig.~\ref{fig:LGA}, the 2$q$VZ is
characterized by $\widetilde{C}(\tau )>0$, i.e., $q_{i}$-susceptibles are
correlated in such a way that $\langle x_{1}(t)x_{2}(t+\tau )\rangle ^{\ast
}>\langle x_{2}(t)x_{1}(t+\tau )\rangle ^{\ast }$. This indicates that on a
finite timescale (up to a separating time $\tau \approx 40-50$ in Fig.~\ref%
{fig:LGA}) $q_{2}$-voters are more likely to \textquotedblleft
follow\textquotedblright\ $q_{1}$-susceptibles than vice-versa. Clearly, 
for large lag times ($\tau \gg \lambda _{\pm }$), $\widetilde{C}
\rightarrow 0$ and correlations are lost.

\subsection{Fluctuation Driven Dynamics below Criticality - Switching Times}
\label{subsec:switchingtimes}

\begin{figure}[tbp]
\centering
\includegraphics[width=\linewidth]{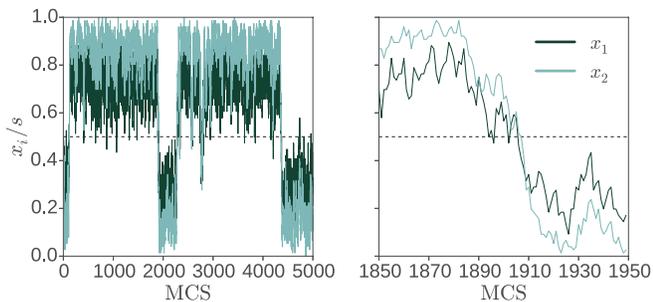}
\caption{\textit{(Color online)}. Switching behavior of the $2q$VZ at $%
z<z_c $: Time series for the total density of $+1$ susceptible voters within
each population ($x_i/s_i$). Sample of $5000$ MCS (left) and closer
examination of a typical switching period (right). One unit of time $t=%
\protect\tau/N$ is $1$ MCS and equates to $N$ iterations of the model. Here
parameters are: $(N,s_1, s_2,z,q_1,q_2)=(200,0.38, 0.38,0.12,1,2)$.}
\label{fig:timeseries}
\end{figure}

Similarly to what happens in the $q$VMZ~\cite{mobilia2015nonlinear} (see also~\cite{nyczka2012opinion}),
the long-time behavior in the low zealotry phase is characterized by a
``swing-state dynamics'', or ``switching dynamics'' when there is the same
number of zealots of both types or when there is only a small asymmetry in
the zealotry. In the switching dynamics both the $q_1$- and $q_2$%
-susceptibles suddenly switch `almost' simultaneously between the peaks $%
x_i^{(\pm)}\approx 0,s$ of $P^*(\vec{x})$, see Fig.~\ref{fig:timeseries}. As
explained above, $q_2$-voters ``follow'' $q_1$-voters and therefore $q_1$%
-susceptibles switch before $q_2$-susceptibles, however the lag between the
switching of both populations is negligible in a first approximation, as shown in Fig.~\ref%
{fig:timeseries} (right), and will be neglected in what follows. In the remainder of this subsection, we focus
on the case of symmetric zealotry, $z_{\pm}=z$.

This switching phenomenon that is driven by fluctuations, and therefore is beyond the
reach of the mean-field analysis, can be analyzed in terms of the {\it mean
switching time} $\tau_s$,
measures the mean time to switch from one peak of $P^*(\vec{x})$, say $%
(x_1^{(-)},x_2^{(-)})$, to the other, $(x_1^{(+)},x_2^{(+)})$ for the first
time~\cite{mobilia2015nonlinear}.

Finding the mean switching time can be formulated as a first-passage problem
whose solution, in terms of the FPE or the Kramer's escape theory~\cite{eyring1935activated,*kramers1940brownian,*hanggi1990reaction}, is
simple only in the single-variate case of a single type of
susceptibles as in the $q$VMZ~\cite{mobilia2015nonlinear}. However, as here the 2$q$VZ
violates the detailed balance, the problem is much more complicated. 

\begin{figure}[tbp]
\centering
\includegraphics[width=0.9\linewidth]{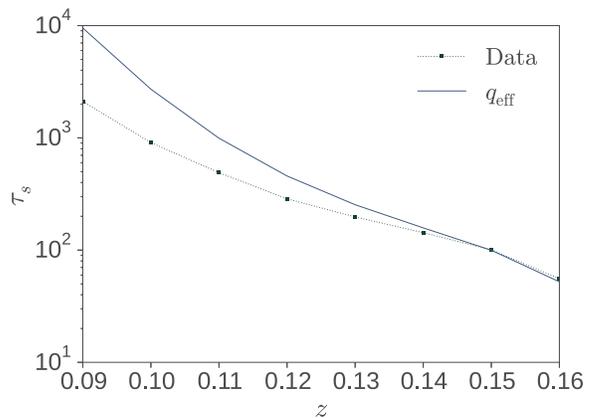} 
\caption{\textit{(Color online)}. Mean switching time $\protect\tau_s$
between the two stable fixed points for systems of size $N=100$ with $(q_1,
q_2)=(1,2)$. 
Results obtained from stochastic simulations (green/black $\square$ with dotted line) are compared with the approximation \protect\ref{approx1} (blue/grey solid line); see text.
Simulation data are averaged over $10^6$ MCS. Here $s_{1,2}=s=\frac{1}{2} - z$.}
\label{fig:switching_times}
\end{figure}

An approximation of $\tau_s$ is obtained by exploiting the mapping onto
the $q$VMZ for values of $z$ just below the critical value $z_c$ ($z\lesssim
z_c$). In this regime, we expect that the switching dynamics of the 2$q$VZ
with $(q_1, q_2)$ can be approximately mapped onto that of the $q$VMZ with an effective
value $q_{\mathrm{eff}}$ given by (\ref{eqn:q_eff}). In this approximation 
the mean switching time between the two fixed points of the qVMZ, $x^{(\pm)}$, can be
computed as in in Ref.~\cite{mobilia2015nonlinear} by solving $\mathcal{G}_b^{\mathrm{eff}%
}(x)\tau_s(x)=-1$ with the reflective and absorbing conditions $(d/d
x)\tau_s(x^{(-)})=\tau_s(x^{(+)})=0$, which yields 
\begin{eqnarray}  \label{approx1}
\tau_s = 2N \int_{x^{(-)}}^{x^{(+)}} dy e^{-N\phi(y)}
\int_{x^{(-)}}^{y} \frac{e^{N\phi(v)}dv}{\tilde{w}^+(v) + \tilde{w}^-(v)},
\end{eqnarray}
where $\phi(v) = 2\int_{x^{(-)}}^{v} \left\{\frac{\tilde{w}^+(v) - \tilde{w%
}^-(v)}{\tilde{w}^+(v) + \tilde{w}^-(v)} \right\}$.
Now, $\tilde{w}^+(x)=(s_1+s_2-x)(x+z)^{q_{\mathrm{eff}}}$ and $\tilde{w}^-(x)=
x(s_1+s_2+z-x)^{q_{\mathrm{eff}}}$ and therefore, using the Kramers' formula~%
\cite{eyring1935activated,*kramers1940brownian,*hanggi1990reaction,mobilia2015nonlinear}, we have 
\begin{eqnarray}  \label{approx2}
\ln{\tau_s} \simeq 2N \int_{x^{(-)}}^{x_1^{(+)}} \frac{\tilde{w}^-(y) - 
\tilde{w}^+(y)} {\tilde{w}^-(x^{(-)}) + \tilde{w}^+(x^{(-)})}~dy
\end{eqnarray}
which predicts that the mean switching time $\tau_s$ grows (approximately) exponentially with $N$~\cite{mobilia2015nonlinear}
The results reported in Fig.~\ref{fig:switching_times} confirm that this
approximation is accurate just below $z_c$, while it overestimates $\tau_s$
at lower values of $z$. The fact that the mean switching time in the 2$q$VZ is generally shorter than in its 
equilibrium $q$VMZ (with $q=q_{\mathrm{eff}}$) counterpart 
suggests that the probability currents, absent in the $q$VMZ, are responsible for speeding up the switching dynamics by reducing the switching time.

\subsection{Behavior Near Criticality}

Up to now, we have mostly focused on the properties on the 2$q$VZ deep in
the high-zealotry and low-zealotry phases. Here we focus on some properties
of the system close to criticality: $z_{\pm }=z\approx z_{c}$.

\begin{figure}[tbp]
\centering
\includegraphics[width=0.9\linewidth]{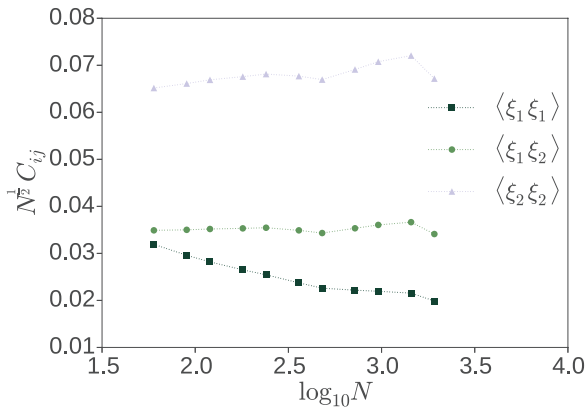}
\caption{\textit{(Color online)}. Correlations as a function of system size $%
\log_{10}(N)$ for simulated results, averaged over at least $10^5$ MCS. The
fluctuations roughly scale as $N^{-1/2}$ as predicted. Here the parameters
are: $(s_1, s_2, z, q_1, q_2)=(\frac{1}{3}, \frac{1}{3}, \frac{1}{6},1,2)$. $N$ varies from $N=60$
to $1920$.
}
\label{fig:critical_correlations}
\end{figure}

Since, every individual in our system is connected to every other (i.e., a
complete graph), we anticipate that a Landau-like description should be
adequate to capture the critical behavior. Specifically, we are concerned
with how the fluctuations (co-variances $C_{ij}$) scale with $N$ at
criticality. By formal analogy with the standard approach for an Ising
magnet, we consider a free energy functional (at zero external magnetic
field) \cite{goldenfeld1992lectures,*tauber2014critical}.
\begin{equation*}
\mathcal{F}\left( m\right) =N\left\{ \frac{rm^{2}}{2}+\frac{gm^{4}}{4}%
\right\} 
\end{equation*}%
where $r$ is a measure of the deviation from the critical point (e.g., $%
T-T_{c}$) and $g>0$. Thus, using $P\left( m\right) \propto e^{-\mathcal{F%
}}$, we find $\left\langle m^{2}\right\rangle \propto N^{-1}$ and $N^{0}$,
above and below criticality respectively - as reported in the previous
section. For $r=0$, however, $\left\langle m^{2}\right\rangle \propto
N^{-1/2}$. In Fig.~\ref{fig:critical_correlations}, we report that in line
with the above general considerations all three stationary two-point
correlation functions $\langle \xi _{i}\xi _{j}\rangle $ of the 2$q$VZ
measured from simulations with $z=z_{c}$ indeed approach this behavior as $N$
increases: $\langle \xi _{i}\xi _{j}\rangle \sim N^{-1/2}$.

A full finite-size-scaling analysis of the entire critical region should
include (i) the behavior of $z_{c}$ as a function of $N$, (ii) the effects
of $z_{+}\neq z_{-}$, (iii) the Binder cumulant (excess kurtosis), etc. We
fully expect universal behavior, i.e., properties which do not depend on the
detailed partition into the two populations ($s_{1},s_{2}$). While valuable,
such a study is beyond the scope of this paper.

\section{The 2$q$VZ with Asymmetric Zealotry}

\label{sec:asymmetric} In this section we briefly outline the main
properties of the 2$q$VZ with asymmetric zealotry by assuming that, say, $%
Z_+ > Z_-$.

\begin{figure}[tbp]
\centering
\includegraphics[width=\linewidth]{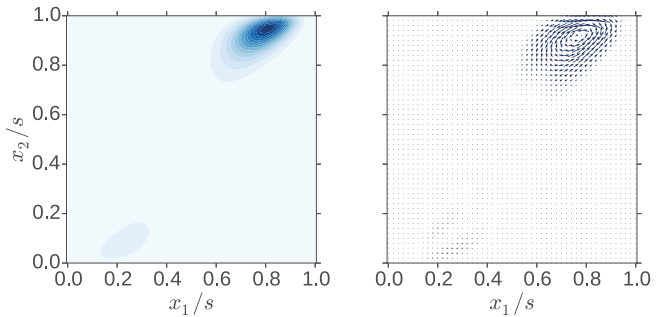} 
\caption{\textit{(Color online)}. Numerically exact NESS probability distribution $P^*$
and probability current $\vec{K}^*$ in the low-zealotry phase with
asymmetric zealotry: (Left) Darkness indicate regions in the $x_1/s-x_2/s$
where $P^*$ is high ($P^*$ being higher where it is darker). (Right) Vector
field of the probability current $\vec{K}^*$. Parameters are: $(N,S_1, S_2,Z_+,
Z_-,q_1,q_2)=(258,100, 100,30, 28,1,2)$.}
\label{fig:AsymZ}
\end{figure}

The 2$q$VZ with asymmetric zealotry shares qualitatively the same features
as its $q$VMZ counterpart~\cite{mobilia2015nonlinear}: It also displays a high- and
low-zealotry phase which are respectively characterized by a unimodal and
bimodal stationary probability distribution $P^*$. However, the latter are
now no longer symmetric: in the low-zealotry phase the peak associated with
the opinion supported by $Z_+$ zealots is much more pronounced than that
associated with $Z_-$ zealots (see Fig.~\ref{fig:AsymZ}(left)), whereas the
single-peaked distribution in the high-zealotry phase is skewed towards
states with a majority of $+1$ voters. When $N$ is sufficiently large, the
peaks of $P^*$ of course correspond to the fixed points of the mean-field
equation (\ref{eqn:rate_equation}) whose analysis reveals that, as for the $%
q $VMZ, in the low-zealotry phase the two stable fixed points $\vec{x}%
^{(\pm)} $ are separated by the unstable steady state $\vec{x}^{(0)}$, while
in the high-zealotry phase only $\vec{x}^{(+)}$ is stable. However, again, a
major difference between the 2$q$VZ and the $q$VMZ is that the former
violates the detailed balance which results in the steady state being a NESS
and in a number of intriguing results such as the existence of stationary
current of probability that form closed loops in the configuration space. A
typical example of the exact NESS probability distribution $P^*$ in the low
zealotry phase with $Z_+>Z_-$ is shown in Fig.~\ref{fig:AsymZ} along with
the wind field of the stationary probability current $\vec{K}^*$. 
In Fig.~\ref{fig:AsymZ}(left), we see that even when there is a small
asymmetry in the zealotry the peak corresponding to $\vec{x}^{(-)}$ is
almost invisible and the distribution is dominated by the peak associated
with $\vec{x}^{(+)}$ and whose intensity increases with $N$. As a
consequence, the stationary probability current $\vec{K}^*$ flows mostly
around $\vec{x}^{(+)}$ in an anti-clockwise fashion (vorticity is positive),
even though there is current flow around $\vec{x}^{(-)}$ but with a much
smaller amplitude. These features of the 2$q$VZ with asymmetric zealotry can
be analyzed using the techniques of Sec.~IV and more specifically the
Fokker-Planck and linear Gaussian approximations in the continuum limit when 
$N\gg 1$. In fact, we can still use (\ref{eqn:LGA2})-(\ref{eqn:C-tilde}) within the LGA
and obtain the stationary probability density $P^*(\vec{\xi})$ near the fixed
points in each of the phase, as well as the LGA expressions of the
stationary density current $\vec{K}^*(\vec{\xi})$, correlation functions $
C_{ij}=\langle \xi_i \xi_j\rangle$ and $\widetilde{C}(t)$, vorticity and
average probability angular momentum $\langle \mathcal{L} \rangle^\ast$. Since
the bimodal probability density $P^*(\vec{\xi})$ is dominated by the peak $
\vec{x}^{(+)}$ in the low-zealotry phase, the LGA is expected to work even
better when the zealotry is asymmetric since the skewness that characterizes 
$P^*(\vec{\xi})$ in this regime disappears when $N$ is large enough.

\begin{figure}[tbp]
\centering
\includegraphics[width=0.9\linewidth]{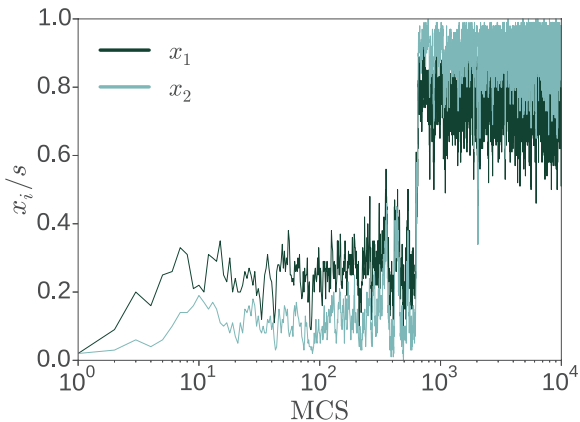}
\caption{\textit{(Color online)}. Metastability and switching dynamics in
the low-zealotry phase with asymmetric zealotry: Shown is the first $10^4$
MCS of a typical single realization with initial condition $\vec{x}=(0,0)$
and switching occurring around $t\approx 1000$. Parameters are: $(N,S_1, S_2,Z_+,
Z_-,q_1,q_2)=(258,100,100,30, 28,1,2)$.}
\label{fig:meta}
\end{figure}

When the zealotry is asymmetric the long-time dynamics is characterized by
metastability: While in principle chance fluctuations can cause the
continuous switching of susceptibles population from the (metastable state) $%
\vec{x}^{(-)}$ into the state $\vec{x}^{(+)}$ and vice versa, owed to the
asymmetry in $P^*$ only the switch from $\vec{x}^{(-)}$ to $\vec{x}^{(+)}$
is observable in practice (when $Z_+ > Z_-$), as in the $q$VMZ~\cite{mobilia2015nonlinear}.
This phenomenon is associated with the metastability of the fixed points $%
\vec{x}^{(\pm)}$ and as such is triggered by a rare large fluctuation: As
shown in Fig.~\ref{fig:meta}, a chance fluctuation causes the sudden and
almost simultaneous switch of both $q_i$-subpopulations from $\vec{x}^{(-)}$
to $\vec{x}^{(+)}$. In a large population, the system settles in the new
metastable and fluctuates in its vicinity but cannot be observed to switch
back to $\vec{x}^{(-)}$. The mean time for the transition $\vec{x}^{(-)}$ to 
$\vec{x}^{(+)}$ can be estimated by using the same approach as in the case of symmetric zealotry
and also gives a mean time that grows exponentially with the population
size, which is typical of a system characterized by metastability.

\section{Discussion and Conclusion}

\label{sec:conclusion} This work has been dedicated to the detailed analysis
of a heterogeneous out-of-equilibrium nonlinear voter model in a finite
well-mixed population. More specifically, we have carefully analyzed the
properties of 2$q$VZ model that we introduced in Ref.~\cite{mellor2016characterization}
where two types of swing voters need the consensus of $q_{1}$ or $q_{2}(\neq
q_1)$ of their neighbors to adopt their opinion and are influenced by the
presence of zealots. The 2$q$VZ is a direct but non-trivial generalization
of the $q$-voter model with zealots ($q$VMZ)~\cite{mobilia2015nonlinear}. Both models mimic
important concepts of social psychology and sociology, such as the relevance
group-size on the mechanism of conformity for collective actions, and the
interplay between independence and conformity. However, from a mathematical
viewpoint the 2$q$VZ strikingly differs from the $q$VMZ by violating the
detailed balance and is therefore genuinely \textit{out of equilibrium}.
Many of the qualitative features of the 2$q$VZ are similar to those
displayed by the $q$VMZ: When the zealotry density is low, the long-time
opinion distribution is bimodal whereas it is single-peaked at high
zealotry. Furthermore, the 2$q$VZ is also characterized by a switching
dynamics when there is an equal number of zealots of both types and by
metastability when there is an asymmetry in the zealotry. However, the
non-equilibrium nature of the 2$q$VZ has far-reaching consequences that we
have investigated in detail. In particular, we have characterized the
system's non-equilibrium steady state (NESS) in terms of its probability
distribution and probability currents that form closed loops in the state
space, as well as the unequal-time correlation functions to highlight the
violation of the time-reversal. These quantities have been computed
numerically exactly for small systems and by means of stochastic simulations
in larger populations. Furthermore, we have also investigated these
quantities analytically in the realm of the linear Gaussian approximation
(LGA) obtained in the limit of large systems and far from the criticality.
We have also focused on the long-time switching dynamics of the 2$q$VZ in
the low-zealotry phase and have devised an approximation 
by mapping the mean switching time of the 2$q$VZ on that computed in Ref.~\cite{mobilia2015nonlinear} for the $q$VMZ. We have also studied some properties of these systems
at criticality and outlined the behavior of the 2$q$VZ when the zealotry is
asymmetric.

This work has allowed us to draw a comprehensive picture of the various
properties of the 2$q$VZ and of the consequences of the violation of the
detailed balance and we have shown that LGA is a particularly useful method.
It shall be interesting to investigate this model on complex and/or adaptive
networks that would be relevant from a social dynamics perspective. An
intriguing question is the macroscopic interpretation of the closed loops of
probability current: can these be related to the presence of ``leaders'' and
``followers'' in a society?

\section{Acknowledgements}

We are grateful to J. Knebel for a critical reading of the manuscript and useful suggestions. 
We thank K. E. Bassler, B. Schmittmann and J. B. Weiss for enlightening discussions.
This work was supported by an EPSRC Industrial Case Studentship Grant No.
EP/L50550X/1. Partial funding from Bloom Agency in Leeds U.K. is also
gratefully acknowledged. This research was supported partly by US National
Science Foundation Grants No. OCE-1245944 and No. DMR-1507371. One of us (R.K.P.Z.) is
grateful for the hospitality of the Leeds School of Mathematics, as well as
financial support from the LSM and the London Mathematical Society (Grant No. 41517). M.M. is grateful 
for the hospitality of the LSFrey at the Arnold
Sommerfeld Center (University of Munich), as well as for the financial 
support of the Alexander von Humboldt Foundation (Grant No. GBR/1119205).

\appendix

\section{Master Equation and Probability Currents}

\label{app:AppendixA} In this appendix, we outline how the probability
current is obtained from the master equation (ME) characterizing the
evolution of a generic interacting stochastic process. For simplicity, let
us focus on a system with configuration space $\left\{ \mathcal{C}\right\} $%
, evolving in discrete time steps. Suppressing the reference to the initial
configuration, $\mathcal{C}_{0}$, we consider the evolution of $P\left( 
\mathcal{C},t\right) $, the probability to find the system in the
configuration $\mathcal{C}$ at time $t$. In a general Markov chain with finite  state space, this
evolution is specified by only the probabilities for the system to
transition from configuration $\mathcal{C}$ to $\mathcal{C}^{\prime }$ in
one time step: $W\left( \mathcal{C}\rightarrow \mathcal{C}^{\prime }\right) $%
. The ME for $P\left( \mathcal{C},t\right) $ can be written as 
\begin{equation}
P\left( \mathcal{C}^{\prime },t+1\right) =\sum_{\mathcal{C}}\mathcal{G}%
\left( \mathcal{C}^{\prime },\mathcal{C}\right) P\left( \mathcal{C},t\right)
,  \label{ProbDist}
\end{equation}%
where the evolution operator $\mathcal{G}$ thus reads 
\begin{equation}
\mathcal{G}\left( \mathcal{C}^{\prime },\mathcal{C}\right) =\delta \left( 
\mathcal{C}^{\prime },\mathcal{C}\right) \left[ 1-\sum_{\mathcal{C}^{\prime
\prime }}W\left( \mathcal{C\rightarrow C}^{\prime \prime }\right) \right]
+W\left( \mathcal{C}\rightarrow \mathcal{C}^{\prime }\right) ,  \label{evolG}
\end{equation}%
where $\delta $ is the Kronecker delta. Clearly, we can regard $P$ as a
vector and $\mathcal{G}$ as a (stochastic) matrix, generating a new vector
at each time step. The \textit{change} in $P\left( \mathcal{C},t\right) $
can be regarded as a sum over\textit{\ }probability \textit{currents} into
and out-of $\mathcal{C}$. Specifically, the \textit{net current} from $%
\mathcal{C}$ to $\mathcal{C}^{\prime }$ is given by 
\begin{equation}
K\left( \mathcal{C}\rightarrow \mathcal{C}^{\prime },t\right) =W\left( 
\mathcal{C}\rightarrow \mathcal{C}^{\prime }\right) P\left( \mathcal{C}%
,t\right) -W\left( \mathcal{C}^{\prime }\rightarrow \mathcal{C}\right)
P\left( \mathcal{C}^{\prime },t\right)  \label{curr}
\end{equation}%
so that%
\begin{equation}
P\left( \mathcal{C},t+1\right) -P\left( \mathcal{C},t\right) =-\sum_{%
\mathcal{C}^{\prime }}K\left( \mathcal{C}\rightarrow \mathcal{C}^{\prime
},t\right)
\end{equation}%
takes the form of a discrete continuity equation. Much is know about
stochastic matrices of the form $\mathcal{G}$. In particular, its largest
eigenvalue is unity and there is (at least) one associated eigenvector,
which can be identified with the stationary state, $P^{\ast }\left( \mathcal{%
C}\right) $. Further, if the dynamics is ergodic (all $\mathcal{C}$'s can be
reached from any $\mathcal{C}_{0}$ with the $W$'s), then $P^{\ast }$ is
unique.

Beyond probabilities at one time, we can consider joint distributions, with
the system evolving from $t$ to a later time $t+\tau $: 
\begin{equation}
\mathcal{P}\left( \mathcal{C}^{\prime },t+\tau ;\mathcal{C},t\right) =%
\mathcal{G}^{\tau }\left( \mathcal{C}^{\prime },\mathcal{C}\right) P\left( 
\mathcal{C},t\right)  \label{jointP}
\end{equation}%
where $\mathcal{G}^{\tau }$\ is the matrix product of $\mathcal{G}$\ with
itself, $\tau $ times.

If detailed balance is satisfied (i.e., if $W$'s satisfy the Kolmogorov
criterion~\cite{kolmogorov1936zur}), then all currents in the stationary state  
\begin{equation}
K^{\ast }\left( \mathcal{C}\rightarrow \mathcal{C}^{\prime }\right) =W\left( 
\mathcal{C}\rightarrow \mathcal{C}^{\prime }\right) P^{\ast }\left( \mathcal{%
C}\right) -W\left( \mathcal{C}^{\prime }\rightarrow \mathcal{C}\right)
P^{\ast }\left( \mathcal{C}^{\prime }\right)  \label{kol}
\end{equation}%
vanish, yielding $K^{\ast }=0$. When detailed balance is violated, then the stationary state is a
NESS and some $K^{\ast }$ will be non-vanishing. Stationarity implies that
currents $K^{\ast }$ are \textquotedblleft divergence\textquotedblright\
free, i.e., $\sum_{\mathcal{C}^{\prime }}K^{\ast }\left( \mathcal{C}%
\rightarrow \mathcal{C}^{\prime }\right) =0$. This means that $K^{\ast }$
must be the `curl' of a certain quantity, while its `curl' generally does
not vanish. In other words, the currents form \textit{closed loops} in $%
\mathcal{C}$\ space. By analogy with the current $\vec{J}$ in classical
mechanics of incompressible fluids (where $\vec{\nabla}\cdot \vec{J}=0$),
these current loops lead us to other concepts, notably the vorticity field ($%
\vec{\omega}=\vec{\nabla}\times \vec{J}$), the stream function ($\vec{J}=%
\vec{\nabla}\times \vec{\psi}$), and total angular momentum ($\int_{\vec{r}}%
\vec{r}\times \vec{J}$).

\subsection{Probability Current and Observables in the 2$q$VZ}

\label{app:AppendixA2}

Here, we study discrete version of these quantities in a very simple
configuration space: In the 2$q$VZ, the $\mathcal{C}$\ space is just the set
of integer points $\left( n_{1},n_{2}\right) \in S_{1}\times S_{2}$. With (%
\ref{W}), the evolution operator of the 2$q$VZ is given by (\ref{Gexpl}).
Hence, with (\ref{curr}) and (\ref{W}), the probability net current
associated in the 2$q$VZ is given by 
\begin{eqnarray}
K_{1}(\vec{n};t) &=&W_{1}^{+}(\vec{n})P\left( \vec{n};t\right)  \notag
\label{K12} \\
&-&W_{1}^{-}(\vec{n}+\vec{e}_{1})P\left(\vec{n}+\vec{e}_{1};t\right)  \notag \\
K_{2}(\vec{n};t) &=&W_{2}^{+}(\vec{n})P\left( \vec{n};t\right)  \notag \\
&-&W_{2}^{-}(\vec{n}+\vec{e}_{2})P\left(\vec{n}+\vec{e}_{2};t\right) .
\end{eqnarray}%
In the continuum limit, $\vec{n}/N\rightarrow \vec{x}$, the usual notation
of divergence and curl applies. Meanwhile, $\vec{K}^{\ast }\left( \vec{x}%
\right) =(K_{1}^{\ast }\left( \vec{x}\right) ,K_{2}^{\ast }\left( \vec{x}%
\right) )$ is a two component vector field while both the vorticity and
stream function are scalar fields: 
\begin{equation}
\omega ^{\ast }\left( \vec{x}\right) =\varepsilon _{ij}\frac{\partial
K_{j}^{\ast }\left( \vec{x}\right) }{\partial x_{i}};~~K_{i}^{\ast }\left( 
\vec{x}\right) =\varepsilon _{ij}\frac{\partial \psi \left( \vec{x}\right) }{%
\partial x_{j}}  \label{vor-stream}
\end{equation}%
where $\varepsilon _{ij}$ is the two-dimensional Levi-Civita symbol and
repeated indices are summed. The interpretation of these quantities is the
following: Vorticity $\omega ^{\ast }$ represents the `essence' or `source'
of $\vec{K}^{\ast }$, much like electric currents are the sources of
magnetic field. On the other hand, $\vec{K}^{\ast }$ is just a $90^{\circ }$
rotation of $\vec{\nabla}\psi ^{\ast }$ and $\psi ^{\ast }$ therefore
provides a simple way to visualize the probability current.

Since it is generally difficult to visualize probability currents, it is
also useful to introduce a quantity $L_{ij}$, which is the (average)
``probability angular momentum''~\cite{shkarayev2014exact,mellor2016characterization} and the formal analog
of the total angular momentum of a fluid: 
\begin{equation}
L_{ij} =\int_{\vec{x}}\varepsilon _{ij}x_{i}K_{j}^{\ast }(\vec{x})~d{\vec{x}}
= \int_{\vec{x}} \left[x_{i}K_{j}^{\ast }(\vec{x}) - x_{j}K_{i}^{\ast }(\vec{%
x})\right]~d{\vec{x}},  \label{PAM}
\end{equation}
where $i,j\in(1,2)$. The probability angular momentum has a single
independent component, say $L_{12} =-L_{21}=\langle \mathcal{L} \rangle^\ast$.
Since, $\vec{K}^{\ast }$ is divergence-free ($\partial K_{i}/\partial
x_{i}=0 $), $\langle \mathcal{L} \rangle^\ast$ is independent of the choice of
the origin of $\left\{ \vec{x}\right\} $.

We are particularly interested in the average number of $q_i$-susceptibles
and their two-point correlation function in the NESS of the 2$q$VZ, which
read: 
\begin{eqnarray}  \label{correl}
\langle n_i \rangle^*&=& \sum_{\vec{n}}n_i P^{\ast}(\vec{n}),  \notag \\
\langle n_i^{\prime }n_j\rangle^*_T&=&\sum_{\vec{n},\vec{n}^{\prime }}
n_i^{\prime }n_j \mathcal{P}^{*}(\vec{n},T; \vec{n}^{\prime },0).
\end{eqnarray}
In particular, the lagged correlation $C_{ij}(\tau)=\langle x_i^{\prime
}x_j\rangle_{N\tau}=\langle n_i^{\prime }n_j\rangle_T/N^2$, is directly related to 
$\langle \mathcal{L} \rangle^\ast$ (\ref{PAM}) when $T=1$. In fact, 
in continuous time $\tau \to t=T/N$, the average probability
angular momentum $\langle \mathcal{L}\rangle^\ast$ is the antisymmetric part $\widetilde{C}_{ij}=C_{ij}-C_{ji}$ of
this two-point unequal-time correlation function in the NESS~\cite{mellor2016characterization}: $\langle \mathcal{L}\rangle^* =
\left. \partial_\tau \widetilde{C}(\tau)\right\vert_0$; see Eq.~(\ref{C+L}). %

\bigskip

\bigskip

The proof for $\left\langle \mathcal{L}\right\rangle^\ast =\left\langle \vec{n}
\times \left( \vec{W}^{+}-\vec{W}^{-}\right) \right\rangle $ in NESS starts
with%
\begin{equation*}
\left\langle \mathcal{L}\right\rangle^\ast =\sum_{\vec{n}}n_{1}K_{2}^{\ast
}\left( \vec{n}\right) -n_{2}K_{1}^{\ast }\left( \vec{n}\right)
\end{equation*}%
and substituting in the expression for stationary flows (\ref{eqn:Kast}). Rearranging various terms on the RHS we
have
\begin{eqnarray}
&=&\phantom{+}\sum_{\vec{n}} \left[ n_{1}W_{2}^{+}(\vec{n})-n_{2}W_{1}^{+}(\vec{n})\right] P^{\ast
}(\vec{n}) \nonumber \\
&&-\sum_{\vec{n}} n_{1}W_{2}^{-}(\vec{n} + \vec{e}_2)P^{\ast }(\vec{n} + \vec{e}_2)\nonumber \\
&&+\sum_{\vec{n}} n_{2}W_{1}^{-}(\vec{n} + \vec{e}_1)P^{\ast }(\vec{n} + \vec{e}_1) \nonumber\\
&=&\phantom{+}\sum_{n_{1},n_{2}=0}n_{1}W_{2}^{+}(n_{1},n_{2})P^{\ast
}(n_{1},n_{2}) \nonumber\\
&&-\sum_{n_{1},n_{2}=0}n_{2}W_{1}^{+}(n_{1},n_{2})P^{\ast
}(n_{1},n_{2}) \nonumber\\
&&+ \sum_{n_{1}=1,n_{2}=0}n_{2}W_{1}^{-}(n_{1},n_{2})P^{\ast
}(n_{1},n_{2}) \nonumber\\
&&- \sum_{n_{1}=0,n_{2}=1}n_{1}W_{2}^{-}(n_{1},n_{2})P^{\ast
}(n_{1},n_{2})  \nonumber\\
\end{eqnarray}
where $\sum_{n_1,n_2=0}$ is the sum over both indices, starting at $0$.
For the last two terms, we can sum over $n_{i}^{\prime }=n_{i}+1$ starting
at $n_{i}^{\prime }=1$. But, since 
\begin{equation*}
W_{1}^{-}(0,n_{2})=0=W_{2}^{-}(n_{1},0)
\end{equation*}%
this sum can be extended to $n_{i}^{\prime }\in \left[ 0,S_{i}\right] $.
\begin{eqnarray}
&=&\phantom{+}\sum_{n_{1},n_{2}=0}\left[
n_{1}W_{2}^{+}(n_{1},n_{2})-n_{2}W_{1}^{+}(n_{1},n_{2})\right] P^{\ast
}(n_{1},n_{2}) \nonumber\\
&&+ \sum_{n_{1},n_{2}=0}\left[ n_{2}W_{1}^{-}(n_{1},n_{2})-n_{1}W_{2}^{-}(n_{1},n_{2})\right] P^{\ast }(n_{1},n_{2}) \nonumber
\end{eqnarray}%
Combining everything, we have%
\begin{equation*}
\left\langle \mathcal{L}\right\rangle^\ast = \sum_{\vec{n}}\left[ n_{1}\left( W_{2}^{+}-W_{2}^{-}\right) -n_{2}\left(
W_{1}^{+}-W_{1}^{-}\right) \right] P^{\ast }.
\end{equation*}
\bigskip

\bigskip

\bigskip

\bigskip

\bigskip

\section{Calculating $P^*$ for large systems}

\label{app:numerics}

For small systems, the full master equation is solvable numerically to find
the stationary state $P^*(\vec{n})$, and subsequently the stationary currents 
$\vec{K}^*$. This is very useful for assisting in the understanding of the
model. However this approach is problematic in that the number of non-zero entries in powers of the transition matrix
$\mathcal{G}=[{\cal G}(\vec{n'},\vec{n})]$
grows approximately quadratically with the power. As there are $(S_1+1) \times
(S_2+1)$ possible states, the transition matrix has $(S_1+1)^2(S_2+1)^2$
entries. For a systems with $S = (S_1+S_2)/2 > 100$, storing these
transition matrices fully in memory becomes non-trivial. For $S=300$, the
required memory is $65$GB (with double precision) which is far beyond the
realms of standard computer setups.

Thankfully it is possible to attain the stationary distribution for larger
systems relatively quickly, managing a trade-off between computational
intensity and memory. After computing the stationary distribution of a
smaller system exactly we use linear interpolation to approximate the
stationary distribution, $\tilde{P}^{\ast }(\vec{x})$, of a system which is
a factor $r$ larger (or a factor $r$ more granular when considering the
densities). In particular, for $r=2$, we begin with%
\begin{align*}
\tilde{P}^{\ast }(\vec{x})=P^{\ast }(\vec{x})
\end{align*}
for $x_{i}=n_{i}/N$ (and so $x_{i}=2n_{i}/2N$). 
Then for the \textquotedblleft missing points\textquotedblright\ we use
the averages of their neighbors:%
\[
\tilde{P}^{\ast }\left( \frac{2n_{i}+1}{2N},\frac{2n_{j}}{2N}\right) =\frac{1%
}{2}\sum_{\eta =0,1}P^{\ast }\left( \frac{n_{i}+\eta }{N},\frac{n_{j}}{N}%
\right) 
\]%
for the $i$-direction, and similarly for the $j$-direction. Finally%
\[
\tilde{P}^{\ast }\left( \frac{2n_{i}+1}{2N},\frac{2n_{j}+1}{2N}\right) =%
\frac{1}{4}\sum_{\eta ,\kappa =0,1}P^{\ast }\left( \frac{n_{i}+\eta }{N},%
\frac{n_{j}+\kappa }{N}\right) \text{ }
\]
We then use this approximate $\tilde{P}^{\ast }$ as an initial distribution
for further iteration. Since $\tilde{P}^{\ast }$ closely resembles the
actual stationary distribution, very few iterations are needed to reach good
accuracy. This means that only lower powers of $\mathcal{G}$ need to be
calculated, alleviating possible memory issues for intermediate system sizes.

\section{Calculating $z_c$ and $q_{\normalfont \text{eff}}$ for arbitrary
numbers of species}

\label{app:z_critcal}

In the main text (Sec. \ref{sec:descriptions}) we calculate the critical
zealotry density $z_{c}$ for a system consisting of two subpopulations, $%
q_{1}$- and $q_{2}$-susceptibles. For systems with symmetric zealotry ($%
z_{\pm }=z$) we are able to calculate $z_{c}$ for a system with an arbitrary
number $N_{s}$ of subpopulations with a distribution of $q_{i}$'s ($i=1,2,\dots
,N_{s}$). Starting with the mean-field rate equations that are direct
generalization of (\ref{eqn:rate_equation}) 
\begin{equation}
\frac{d}{dt}x_{i}=w_{i}^{+}-w_{i}^{-}=(s_{i}-x_{i})\mu ^{q_{i}}-x_{i}(1-\mu
)^{q_{i}}.  \label{genRE}
\end{equation}%
where $\mu \equiv z_{+}+\sum_{i=1}^{N_{s}}x_{i}$. %
The fixed points $x_{i}^{\ast }$ of (\ref{genRE}) are given by 
\begin{equation*}
\frac{s_{i}}{x_{i}^{\ast }}=1+\rho ^{q_{i}},
\end{equation*}%
where $\rho \equiv \left( 1-\mu ^{\ast }\right) /\mu ^{\ast }$%
satisfies 
\begin{equation*}
\mu ^{\ast }=z_{+}+\sum_{i-1}^{N_{s}}\frac{s_{i}}{1+\rho ^{q_{i}}}=\frac{1}{%
1+\rho }.
\end{equation*}

We now specialize to the case of symmetric zealotry $z_{\pm }=z$. In this
case, Eqs.~\ref{genRE} have always a fixed point $\vec{x}%
^{(0)}=(s_{1}/2,s_{2}/2,\dots ,s_{N_{s}}/2)$ at the center for which 
\begin{equation*}
\rho =1\quad \text{and}\quad \mu ^{\ast }=1/2.
\end{equation*}%
As in the case $N_{s}=1,2$~\cite{mobilia2015nonlinear,mellor2016characterization}, the generalized model exhibits
criticality if $\vec{x}^{(0)}$ changes stability at some critical zealotry
density $z_{c}$. The method for finding $z_{c}$ follows as in the main text,
exploiting the generalized stability matrix $\mathbb{F}(\vec{x}%
^{(0)})=-[\partial \dot{x_{i}}/\partial x_{j}|_{\vec{x}^{(0)}}]$ 
\begin{eqnarray*}
-\left. \frac{\partial \dot{x_{i}}}{\partial x_{j}}\right\vert _{\vec{x}%
^{(0)}} &=&-\left. \frac{\partial \left[ (s_{i}-x_{i})\mu
^{q_{i}}-x_{i}(1-\mu )^{q_{i}}\right] }{\partial x_{j}}\right\vert _{\vec{x}%
^{(0)}} \\
&=&-\delta _{ij}\left( \frac{1}{2}\right) ^{q_{i}}+\frac{s_{i}}{2}%
q_{i}\left( \frac{1}{2}\right) ^{q_{i}-1} \\
&&-\delta _{ij}\left( \frac{1}{2}\right) ^{q_{i}}-\frac{s_{i}}{2}q_{i}\left( 
\frac{1}{2}\right) ^{q_{i}-1}\left( -1\right)  \\
&=&2^{1-q_{i}}\left[ \delta _{ij}-s_{i}q_{i}\right] 
\end{eqnarray*}%
Now, $\det {\mathbb{F}(\vec{x}^{(0)})}$ is obtained from Sylvester's
determinant theorem:%
\begin{equation*}
\det {\mathbb{F}(\vec{x}^{(0)})}=2^{N_{s}-\Sigma _{i}q_{i}}~\left(
1-\sum_{i=1}^{N_{s}}s_{i}q_{i}\right) 
\end{equation*}%
so that, the criticality condition, $\det \mathbb{F}(\vec{x}^{(0)})=0$,
yields 
\begin{equation}
\sum_{i=1}^{N_{s}}s_{i}q_{i}=1.  \label{eqn:app_critical_z}
\end{equation}%
Hence, the critical zealotry density is $z_{c}=(1-\sum_{i=1}^{N_{s}}s_{i})/2$
where the $s_{i}$ and $q_{i}$'s are subject to satisfy (\ref%
{eqn:app_critical_z}). As a result, it is the same as if we had a
homogeneous population (i.e., $s_{\text{eff}}=\sum_{i=1}^{N_{s}}s_{i}$) with
an effective $q=q_{\text{eff}}$ given by $s_{\text{eff}}q_{\text{eff}}=1$.
Thus, we find%
\begin{equation*}
q_{\text{eff}}=\frac{\sum_{i=1}^{N_{s}}q_{i}s_{i}}{\sum_{i=1}^{N_{s}}s_{i}},
\end{equation*}%
which can easily be interpreted as the \textquotedblleft average $q$%
\textquotedblright\ of the population.

\section{Explicit forms of $\mathbb{C}$ \& $\protect\widetilde{\mathbb{C}}%
(t) $ in the LGA}

\label{app:explicit} In this appendix we show how to compute the correlation
matrix $\mathbb{C}$ used in the realm of the linear Gaussian approximation
(LGA) of Sec. IV and give its explicit expression in the symmetric case ($%
Z_+=Z_-=Z$ and $S_1=S_2=S$) with $q_1=1$ and $q_2=2$. We also outline the
LGA calculation of $\widetilde{\mathbb{C}}(\tau)$,
the antisymmetric part  of
the unequal-time correlation function $C_{ij}(\tau)=\langle \xi_i(\tau) \xi_j(0)
\rangle^*$ in the NESS.

\subsection{Explicit form of $\mathbb{C}$ within the LGA}

As explained in Sec.~IV, in the realm of the LGA the stationary probability
about a fixed point $\vec{x}^{\ast }$ is given by $P^{\ast }(\vec{\xi}%
)\propto \mathrm{exp}\left[ -\frac{1}{2}\vec{\xi}\cdot \mathbb{C}^{-1}\vec{%
\xi}\right] $, where $\vec{\xi}=\vec{x}-\vec{x}^{\ast }$ and $\mathbb{C}%
=[C_{ij}],i,j=1,2$ is the symmetric real correlation matrix. The stationary
probability density current within the LGA is $\vec{K}^{\ast }=-(\mathbb{F}%
\mathbb{C}-\mathbb{D})\mathbb{C}^{-1}\vec{\xi}~P^{\ast }(\vec{\xi})$, where $%
\mathbb{F}$ is the stability matrix. Since $\vec{K}^{\ast }$ is divergence-free, $%
\mathbb{F}\mathbb{C}-\mathbb{D}$ has to be antisymmetric and we thus have~%
\cite{weiss2003coordinate, *weiss2007fluctuation,zia2007probability} 
\begin{equation}
\mathcal{S}(\mathbb{F}\mathbb{C})=\mathbb{D},  \label{eqC}
\end{equation}%
where $\mathcal{S}(\mathbb{F}\mathbb{C})=(\mathbb{F}\mathbb{C}+\mathbb{C}%
\mathbb{F}^{T})/2$ is the symmetric part of $\mathbb{F}\mathbb{C}$. Since
the matrices $\mathbb{F}$ and $\mathbb{D}$ are thus readily obtained as
explained the main text, the expression of the correlation matrix is
obtained by solving (\ref{eqC}). It is useful to remind the reader that in  
the main text, see Eq.~(\ref{eqn:L}), we have shown that 
$\mathbb{L}=[L_{ij}]=2\mathcal{A}(\mathbb{F}
\mathbb{C})$ where $\mathcal{A}(\mathbb{F}\mathbb{C})=(\mathbb{F}\mathbb{C}-
\mathbb{C}\mathbb{F}^{T})/2$ is the antisymmetric part of $\mathbb{F}\mathbb{
C}$ and $\mathbb{F}$ is the stability matrix around $\vec{x}^{\ast }$ (see also Refs.~\cite{shkarayev2014exact,mellor2016characterization}).

In the symmetric case $Z_+=Z_-=Z$ and $S_1=S_2=S$, with $q_1=1$ and $q_2=2$,
the explicit expressions of $\mathbb{F}$ and $\mathbb{D}$ around each fixed
point are given in~\cite{mellor2016supp}. With those quantities, we find the
following expressions of the covariance matrix:

(i) Around the fixed point $\vec{x}^{(0)}$ (with $s<1/3$): 
\begin{eqnarray}
\mathbb{C}^{(0)} &=&\frac{s}{4N(1-3s)(3-4s)}  \label{eqn:C0} \\
&\times &\left( 
\begin{array}{cc}
3-10s+6s^{2} & 2s(2-3s) \\ 
2s(2-3s) & 3-7s+6s^{2}%
\end{array}%
\right) .
\end{eqnarray}%
With this expression and that of the stability matrix $\mathbb{F}^{(0)}$
around $\vec{x}^{(0)}$~\cite{mellor2016supp}, the average probability angular
momentum in the LGA is $L_{12}=\langle \mathcal{L}\rangle^{\ast} =\frac{s^{2}}{%
2N(3-4s)}$. 

(ii) Around the fixed points $\vec{x}^{(\pm )}$ (with $z<z_{c}=1/6$): 
\begin{eqnarray}
C_{11}^{(\pm )} &=&\frac{2z(1-2z)(3-11z-24z^{2}-36z^{3})}{%
N(6z-1)(3+2z)(1+2z)^{2}}  \label{eqn:Cpm} \\
C_{22}^{(\pm )} &=&\frac{6z^{2}(1+2z)}{N(6z-1)(3+2z)} \\
C_{12}^{(\pm )} &=&C_{21}^{(\pm )}=\frac{2z^{2}(12z(1+z)-5)}{%
N(1+2z)(3+2z)(6z-1)}.
\end{eqnarray}%
With the expression of $\mathbb{C}^{(\pm )}$ and the stability matrix $%
\mathbb{F}^{(\pm )}$ around $\vec{x}^{(\pm )}$ (see~\cite{mellor2016supp}), the
probability angular momentum in the LGA is $\langle \mathcal{L}\rangle^{\ast} =\frac{4z^{2}}{N(3+2z)}$.  

\subsection{Calculation of $\protect\widetilde{\mathbb{C}}(t)$ within the LGA%
}

In the realm of the LGA, the covariance matrix $\mathbb{C}(t)$ for a NESS
around a given FP is explicitly given by $\mathbb{C}\mathrm{exp}(-\mathbb{F}%
^{T}t)$, where $\mathbb{C}=\mathbb{C}(0)$ is obtained as described above and 
$\mathbb{F}$ is the stability matrix. To obtain the antisymmetric quantity $%
\widetilde{\mathbb{C}}(t)=e^{-\mathbb{F}^{T}t}\mathbb{C-C}e^{-\mathbb{F}t}$,
we exploit Sylvester's formula to write  
\begin{equation*}
e^{-\mathbb{F}t}=\left( \frac{\lambda _{+}e^{-\lambda _{-}t}-\lambda
_{-}e^{-\lambda _{+}t}}{\lambda _{+}-\lambda _{-}}\right) \mathbb{I}+\left( 
\frac{e^{-\lambda _{+}t}-e^{-\lambda _{-}t}}{\lambda _{+}-\lambda _{-}}%
\right) \mathbb{F},
\end{equation*}%
where $\mathbb{I}$ is the identity matrix and $\lambda _{\pm }=\left( 
\mathrm{Tr}\mathbb{F}\pm \sqrt{(\mathrm{Tr}\mathbb{F})^{2}-4\mathrm{det}%
\mathbb{F}}\right) /2$ are the eigenvalues of $\mathbb{F}$. Thus,%
\begin{equation}
\widetilde{\mathbb{C}}(t)=\left( \frac{e^{-\lambda _{-}t}-e^{-\lambda _{+}t}%
}{\lambda _{+}-\lambda _{-}}\right) \underbrace{[\mathbb{F}\mathbb{C}-%
\mathbb{C}\mathbb{F}^{T}]}_{=\mathbb{L}}.  \label{Ctilde}
\end{equation}%
Since $\widetilde{\mathbb{C}}(t)$ is antisymmetric, it has only one
independent quantity, say $\widetilde{C}_{12}(t)=\widetilde{C}(t)$, and with
the probability angular momentum $\langle \mathcal{L}\rangle^\ast =L_{12}$, we
find explicitly $\widetilde{C}(t)=\langle \mathcal{L}\rangle^\ast \left( \frac{%
e^{-\lambda _{-}t}-e^{-\lambda _{+}t}}{\lambda _{+}-\lambda _{-}}\right) $.
For the symmetric case $S_{1}=S_{2}=S$ and $Z_{+}=Z_{-}=N-2S$ with $q_{1}=1$
and $q_{2}=2$, the eigenvalues $\lambda _{\pm }$ are readily obtained from
the explicit expressions of $\mathbb{F}$ given in~\cite{mellor2016supp}.

\end{document}